\documentclass[10pt,journal,compsoc]{IEEEtran}

\ifCLASSOPTIONcompsoc
  \usepackage[nocompress]{cite}
\else
  \usepackage{cite}
\fi
\ifCLASSINFOpdf
\else
\fi
\usepackage{subfigure}
\usepackage{graphicx}
\usepackage{booktabs} % For formal tables
\usepackage{amsmath}
\usepackage{amssymb}
\usepackage{amsthm}
\usepackage{epstopdf}
\usepackage{color}
\usepackage{multirow}
\usepackage{xspace}
\usepackage{makecell}
\usepackage{pifont}
\usepackage{enumitem}
\usepackage{subfigure}
\usepackage{balance}
\usepackage{caption}
 \usepackage{cite}
\newenvironment{pf}{\textit{Proof Sketch:}}{\hfill$\square$}

\newenvironment{definition}{\textit{Definition:}}

\newenvironment{theorem}{\textit{Theorem:}}

\def\ie{\textit{i.e.}}
\def\eg{\textit{e.g.}}
\def\etc{\textit{etc.}\xspace}
\def\etal{\textit{et~al.}\xspace}

\begin{document}

% \title{Privacy-Preserving Collaborative Learning for Cross-Media Retrieval in the Cloud}
\title{Augmenting Encrypted Search: A Decentralized Service  Realization with Enforced Execution}
\author{Shengshan Hu,
    Chengjun Cai,
  Qian Wang,~\IEEEmembership{Senior~Member,~IEEE}, \\
  Cong Wang,~\IEEEmembership{Senior~Member,~IEEE},
    Minghui Li,
    Zhibo Wang,~\IEEEmembership{Senior~Member,~IEEE},
  and Dengpan Ye
\IEEEcompsocitemizethanks{\IEEEcompsocthanksitem S. Hu is with the
Department of Computer Science, City University of Hong Kong, Hong Kong, and also with
the School of Cyber Science and Engineering, Wuhan University, Wuhan, Hubei 430072, China. E-mail: shengshhu2-c@my.cityu.edu.hk.
\IEEEcompsocthanksitem C. Cai and C. Wang are with the Department of Computer Science, City University of
Hong Kong, Hong Kong. E-mail: chencai-c@my.cityu.edu.hk, congwang@cityu.edu.hk.
%\IEEEcompsocthanksitem Z. Qin  is with the Institute of Cyberspace Research, Zhejiang University. E-mail: qinzhan@zju.edu.cn.
\IEEEcompsocthanksitem Q. Wang, M. Li, Z. Wang, and D. Ye are with the School of Cyber Science and Engineering, Wuhan University, Wuhan, Hubei 430072, China. E-mail:\{qianwang, minghuili, zbwang, yedp\}@whu.edu.cn.
\IEEEcompsocthanksitem  A preliminary version of this work was published in IEEE INFOCOM 2018~\cite{hu2018searching}.
%\IEEEcompsocthanksitem Q. Li is with Graduate School at Shenzhen, Tsinghua University, Shenzhen, China. E-mail: qi.li@sz.tsinghua.edu.cn
%\IEEEcompsocthanksitem K. Ren is with the Department of Computer Science and Engineering, The State University of New York, Buffalo, NY 14260 USA. E-mail:kuiren@buffalo.edu
}% <-this % stops an unwanted space
}

%{Shell \MakeLowercase{\textit{et al.}}: Bare Demo of IEEEtran.cls for Computer Society Journals}



\IEEEtitleabstractindextext{%
\begin{abstract}
Searchable symmetric encryption (SSE) allows the data owner to outsource an encrypted database to a remote server in a private manner while maintaining the ability for selectively search. So far, most existing solutions focus on an honest-but-curious server, while security designs against a malicious server have not drawn enough attention. A few recent works have attempted to construct verifiable SSE that enables the data owner to verify the integrity of search results.
Nevertheless, these verification mechanisms are highly dependent on specific SSE schemes, and fail to support complex queries.  A general verification mechanism is desired that can be applied to all SSE schemes.
In this work, instead of concentrating on a central server, we explore the potential of the smart contract, an emerging blockchain-based decentralized technology, and construct decentralized SSE schemes where the data owner can  receive correct search results with assurance without worrying about potential wrongdoings of a malicious server. We study both public and private blockchain environments and propose two designs with a trade-off between security and efficiency. To better support practical applications, the multi-user setting of SSE is further investigated where the data owner allows authenticated users to search keywords in shared documents. We implement prototypes of our two designs and present experiments and evaluations to demonstrate the practicability of our decentralized SSE schemes.

\end{abstract}

}

\maketitle

\IEEEdisplaynontitleabstractindextext

% \fi
\IEEEpeerreviewmaketitle

\begin{table*}[t]
\newcommand{\tabincell}[2]{\begin{tabular}{@{}#1@{}}#2\end{tabular}}
\centering
\begin{tabular}{|c||c|c|c|c|c|c|c|c|}
\Xhline{1.1pt}
\multirow{2}{*}{Scheme} & \multicolumn{2}{c|}{Search} & \multicolumn{2}{c|}{Update} &\multirow{2}{*}{Decentralized} &\multirow{2}{*}{Soundness} &\multirow{2}{*}{Stateless}  \\
\cline{2-5}
& Time & Commu. & Time & Commu. & & &  \\
\Xhline{1.1pt}
\multicolumn{8}{|c|}{Previous Works}\\
\Xhline{1.1pt}
CJJ'14~\cite{CaJa14} & $\mathcal{O}(d_w)$ & $\mathcal{O}(t_w)$ & $\mathcal{O}(|\textsf{W}_{\textsf{id}}|)$ & $\mathcal{O}(|\textsf{W}_{\textsf{id}}|)$ & \ding{55} & \ding{55} & \ding{55}\\
\hline
SPS'14~\cite{StPa14} & $\mathcal{O}($min$\{d_w+\log N ,t_w\log^3 N \})$ & $\mathcal{O}(t_w+\log N)$ & $\mathcal{O}(\log^2N)$ & $\mathcal{O}(\log N)$ & \ding{55} & \ding{51} & \ding{55}\\
\hline
BFP'16~\cite{BoFo16} & $\mathcal{O}(d_w+\log m)$ & $\mathcal{O}(t_w+\log m)$ & $\mathcal{O}(\log m)$ & $\mathcal{O}(\log m)$ & \ding{55} & \ding{51} & \ding{55}\\
\hline
B'16~\cite{Ra16} & $\mathcal{O}(d_w)$ & $\mathcal{O}(t_w)$ & $\mathcal{O}(|\textsf{W}_{\textsf{id}}|)$ & $\mathcal{O}(|\textsf{W}_{\textsf{id}}|)$ & \ding{55} & \ding{51} & \ding{55}\\
\Xhline{1.1pt}
\multicolumn{8}{|c|}{This Work}\\
\Xhline{1.1pt}
$\Pi$ & $\mathcal{O}(d_w)$ & $\mathcal{O}(t_w)$ & $\mathcal{O}(|\textsf{W}_{\textsf{id}}|)$ & $\mathcal{O}(|\textsf{W}_{\textsf{id}}|)$ & \ding{51} & \ding{51} & \ding{51}\\
\hline
$\Pi^+$ & $\mathcal{O}(d_w)$ & $\mathcal{O}(t_w)$ & $\mathcal{O}(|\textsf{W}_{\textsf{id}}|)$ & $\mathcal{O}(|\textsf{W}_{\textsf{id}}|)$ & \ding{51} & \ding{51} & \ding{51}\\
\Xhline{1.1pt}
\end{tabular}
\caption{Comparison with existing SSE schemes. Complexities of time and communication costs are presented. $N$ is the total number of keyword/document pairs. $m$ is the number of distinct keywords. $d_w$ is the number of times that the keyword $w$ is \textit{historically added} to the database. $t_w$ is the size of returned search result set. $|\textsf{W}_{\textsf{id}}|$ denotes the number of distinct keywords for a given file identifier $\textsf{id}$. Each update is evaluated by adding/deleting a file. We focus on the schemes that are secure against a malicious adversary (\ie, preserving soundness) here.}
\label{tab:related}
\end{table*}

\IEEEraisesectionheading{\section{Introduction}\label{sec:introduction}}

\IEEEPARstart{S}{earchable} symmetric encryption (SSE) has been extensively studied for a long time since it was first introduced in~\cite{SoWa00}. Generally, it allows a data owner to outsource data to an untrusted server in the encrypted form and later search for the records matching a given query. During the entire search process, the private information about the database and the query is well protected from the semi-trusted server.

In most existing works, the remote server is modeled as an honest-but-curious entity~\cite{KaPa12,JaJu13,CuGa11,HuWWQR16,KaMo17,FiVo15,PaKe14,WangHDCLZ18,WangDCCZCH18} who never tries to deviate from the prescribed protocol. In reality, however, a malicious server may return partial answers or even non-matching documents (\eg, due to random failures).
%For instance, a cloud service provider usually faithfully answers clients' request, and tries its best to guarantee the security in order to maintain its reputation, but in some cases
More seriously, any security breach and insider attacker may  illegally gain access to alter the computations performed over the data. This could happen when a successful malware infection (\eg, email attachments, infected P2P media) on one host gives an attacker a high access authority. To address these concerns, security designs against a malicious server are urgently needed to facilitate the wide application of SSE.

Recently, a few works have been devoted to designing verifiable SSE schemes where a data owner is able to verify the integrity of search results.
Nevertheless, their verification techniques (\eg, using MAC~\cite{StPa14} or hash table~\cite{Ra16}) are highly dependent on specific SSE schemes,
%In other words, each verification technique is confined to only one SSE scheme, and cannot be adapted to other SSE schemes.
%Furthermore, existing verification techniques
and for now only support simple query expressions such as single-keyword search.
%As far as we are concerned, there is no general framework that can
How to generically
impose verifiability on the existing abundant SSE schemes that support expressive queries and complex data structures (\eg, boolen query~\cite{CaJa13,FaJa15} or graph data~\cite{MeKa15}) without incurring expensive overheads on the data owner remains unclear.

We observe that the main reason of possible cheating is that the centralized server takes full control of data and executes protocols independently without being supervised. In light of this, we resort to smart contract, a newly emerging decentralized computing paradigm in blockchain where all operations are transparent and reliable. Getting rid of a central server, outsourcing search queries to smart contract yields a correct and immutable result, and requires no further verifications by the data owner. It thoroughly eliminates our misgivings about a malicious adversary as long as the security of blockchain is guaranteed.

To this end, we first study public blockchain, a permissionless environment that everyone can get access to. It
provides an off-the-shelf decentralized platform,  enabling the data owner to directly make use of it.
By utilizing the popular public blockchain environment Ethereum~\cite{Wo14},
we, for the first time, propose a decentralized SSE scheme $\Pi$.
The smart contract running over Ethereum is carefully designed to circumvent various barriers (\eg, gas limitation, gas availability) in Ethereum.
Considering some application scenarios where a set of permissioned service providers (\ie, peer nodes) is available, we further study private blockchain environment running among those service providers, and  propose an alternative decentralized SSE scheme $\Pi^+$ leveraging the popular private blockchain framework Hyperledger~\cite{Hyperledger}. The two proposed designs $\Pi$ and $\Pi^+$ have their own merits, leading to a trade-off between security and efficiency.
 To give an exemplary instantiation, both $\Pi$ and $\Pi^+$ are constructed on classic inverted index based searchable
symmetric encryption schemes~\cite{CaJa14,Ra16}.
We emphasize that our framework
is a general one, and many other SSE solutions  supporting complex expressiveness (\eg, boolean queries) and structured data (\eg, graph) fit for our setting as well and can be altered likewise to have their decentralized counterparts, as explicitly discussed in Section~\ref{sec:general}.

In order to further support practical applications, we investigate the multi-user setting, a more complex scenario of SSE~\cite{CuGa11,JaJu13,KiOk16} where an authorized user is allowed to search files shared by the data owner.
For instance, in a traditional cloud-based picture or file sharing system (\eg, Dropbox), a data owner can upload its pictures or files to the cloud server such that they can be shared among friends or family. In our decentralized setting, instead of using the cloud server, we also aim to provide sharing services through the blockchain network.
We study public and private blockchain environments and show how to enable users to search private database, and impose search control such as adding or revoking users. According to the characteristics of the underlying blockchain platforms, we use a straightforward extension for $\Pi$: letting the data owner search after receiving the user's query. For the private blockchain scheme $\Pi^+$, we %make use of broadcast encryption~\cite{KiOk16} such that
propose an alternative approach that enables
the user to search keywords independently and efficiently without getting any help from the data owner.

In summary, we make the following key contributions:

\begin{itemize}

\item By leveraging the smart contract, we propose two decentralized searchable symmetric encryption (SSE) schemes $\Pi$ and $\Pi^+$, catering to the public and the private blockchain environments respectively, to guarantee that the data owner receives correct search results and has no need to perform verifications in the face of a malicious adversary.

\item We investigate  the multi-user setting for both $\Pi$ and $\Pi^+$ where the authorized users are able to search shared files correctly and privately, and the data owner can add/revoke users flexibly.

\item We implement two prototypes of $\Pi$ and $\Pi^+$. Extensive experiments and evaluations over local simulated network and official test network demonstrate the practicability of designing SSE schemes in a decentralized manner.

\end{itemize}

\section{Related Work}

\textbf{Searchable Symmetric Encryption.} SSE was first introduced in~\cite{SoWa00}. Since then, great efforts have been devoted to developing secure and efficient SSE schemes. More than ten years ago,~\cite{CuGa11} for the first time formally considered leakage and designed a static SSE scheme that is secure against adaptive chosen-keyword attack. As a following work,~\cite{KaPa12} proposed the first dynamic SSE scheme that is also secure against adaptive chosen-keyword attack.

In recent years, most of SSE works focus on supporting more complex structures and queries and improving efficiency with regard to search time and communication cost. One of the most notable examples is~\cite{CaJa13} that proposed the first SSE scheme to support conjunctive queries in sub-linear time. Then~\cite{FaJa15} extended this work to achieve much more complex queries including substring, range, wildcard and phrase queries. Besides,~\cite{FiVo15,PaKe14} showed how to handle boolean formulas, ranges and stemming by using garbled circuits and bloom filters.~\cite{CaJa14} then proposed several optimizations to handle very-large datasets (\eg, tens of billions of record-keyword pairs). Recently,~\cite{KaMo17} proposed the first efficient disjunctive and boolean SSE scheme with the worst-case sub-linear search complexity and optimal communication overhead.  Along another line,~\cite{ChKa10} extended SSE to support arbitrarily-structured data, such as graphs, labeled data or matrices. And a recent work~\cite{MeKa15} presented a graph encryption scheme to support approximate shortest distance queries. All of these works, however, address the security against a semi-honest adversary. They are vulnerable to a malicious server who may return incorrect search results.

\textbf{Verifiable Searchable Symmetric Encryption.} To mitigate a malicious adversary, verifiable SSE schemes have aroused interests in recent years.~\cite{KuOh12} studied this problem and proposed a verifiable SSE scheme that is UC-secure. Then~\cite{StPa14,BoFo16} constructed  dynamic and  more efficient schemes. Based on these results, recently~\cite{Ra16} used trapdoor permutations to construct a very simple forward secure searchable encryption scheme.
To address the limitations of demanding specific SSE constructions,~\cite{ZhuLW18} proposed a generic verifiable scheme by using Merkle Patricia Tree (MPT) and Incremental Hash to create the proof index.
Nevertheless, these works have to impose extra computation cost and storage overhead on a stateful data owner.
% Besides, their verification techniques are design-dependent and only support limited queries like single-keyword search, whether more complex queries such as shortest distance query or multi keyword search~\cite{du2018privacy} are compatible  is not yet clear.
Our preliminary work~\cite{hu2018searching} proposed utilizing smart contracts in Ethereum  to realize a decentralized and reliable SSE scheme, but it suffers from high overheads (\eg, gas and cryptocurrency consumptions, time costs) due to
some inherent characteristics of public blockchains (\eg, PoW-based mining process), and does not fully consider the multi-user setting where adding/revoking users should be supported. We therefore propose a new scheme $\Pi^+$ by making use of private blockchain to improve efficiency.
We further investigate the multi-user setting for $\Pi^+$, and  show how to enable authorized users to search private database. We  propose new secure protocols  to flexibly add and revoke users. Moreover,  several construction variants are proposed to address some security issues and strengthen our designs.
Table~\ref{tab:related} gives a comparison of our work and previous verifiable SSE schemes.

\section{Preliminaries}

In this section, we provide some basic introductions on traditional searchable symmetric encryption (SSE) and the cryptographic tools we will use, and main technologies that support our decentralized design, namely smart contracts in Ethereum and Hyperledger.

\subsection{Searchable Symmetric Encryption}

We follow the formalization of Kamara \etal~\cite{KaPa12} with a slight modification. In our paper, $\lambda$ is defined as the security parameter and $\textsf{negl}(\lambda)$ denotes a negligible function in the security parameter. The set of all binary strings of length $\lambda$ is denoted as $\{0,1\}^{\lambda}$, and the set of all finite binary strings is denoted as $\{0,1\}^*$. We write $x \xleftarrow{\$} X$ to represent an element being sampled uniformly at random from a finite set $X$. The algorithms and protocols are running in  polynomial time.
% in the security parameter $\lambda$.
In particular, adversaries are polynomial-time algorithms.

A database $\textsf{DB} = (\textsf{id}_i, \textsf{W}_i)_{i=1}^{d}$ is a list of identifier/keyword-set pairs where $\textsf{id}_i \in \{0,1\}^{l}$ and $\textsf{W}_i \subseteq \{0,1\}^*$. The set of keywords of the database $\textsf{DB}$ is $\textsf{W} = \cup_{i=1}^d\textsf{W}_i$. The set of documents containing a given keyword $w \in \textsf{W}$ is denoted as $\textsf{DB}(w) = \{\textsf{id}_i | w \in \textsf{W}_i\}$. We will always set $m = |\textsf{W}|$ and $N = \sum_{w \in \textsf{W}}|\textsf{DB}(w)|$ to be the number of distinct keywords and the total number of keyword/document pairs, respectively.

A traditional \textit{dynamic searchable symmetric encryption} scheme $\Pi$ consists of one algorithm $\textsf{Setup}$ and two protocols $\textsf{Search}$ and $\textsf{Update}$ between a data owner and a server.

\begin{itemize}

\item $\textsf{Setup(DB)}$ takes as input a database $\textsf{DB}$ and outputs a tuple $(\textsf{EDB}, K,\delta)$ where $\textsf{EDB}$ is the encrypted database, $K$ is a secret key, and $\delta$ is the data owner's state.

\item $\textsf{Search}(K,\delta,w;\textsf{EDB})$ is an interactive protocol where the data owner takes as input the secret key $K$, its state $\delta$, and a search word $w \in \{0,1\}^*$, and the server takes as input the encrypted database $\textsf{EDB}$. The server outputs a set of identifiers while the data owner has no output.

\item $\textsf{Update}(K,\delta,\textsf{op}, \textsf{id}, \textsf{W}_{\textsf{id}};\textsf{EDB})$ is an interactive  protocol between the data owner with inputs the key $K$, the state $\delta$, an operation $\textsf{op} \in \{\textsf{add},\textsf{del}\}$, a file identifier $\textsf{id}$, and a set $\textsf{W}_{\textsf{id}}$ of distinct keywords, and the server with input $\textsf{EDB}$. These inputs represent the actions of adding a new file with identifier $\textsf{id}$ and deleting the file with identifier $\textsf{id}$.

\end{itemize}

For simplicity, the formalization of SSE here does not model the storage of the actual document payloads. The SSE literature varies on dealing with this issue. In our case where decentralized environment is considered, we can store encrypted documents in any decentralized file systems such as IPFS discussed below.

\textbf{Cryptographic Tools.} In our constructions,
%we will use CPA-secure (Chosen-Plaintext-Attack-secure) private-key encryption scheme. In addition,
we make use of variable-input-length pseudo-random functions (PRFs) which are polynomial-time computable functions that cannot be distinguished from random functions by any probabilistic polynomial-time adversary. Formal definitions of
%CPA-security and
PRFs can be found in~\cite{KaLi14}. Some of our constructions will be analyzed in the random oracle model~\cite{BeRo93}. We use $H$ to denote the random oracle.

\subsection{Gas System in Ethereum}
Gas system is a fantastic feature in Ethereum. It is designed to mitigate Denial-of-Service (DoS) attack on the Ethereum network. Specifically, the contract script is compiled into Ethereum opcodes and stored in the blockchain. Each opcode will cost a certain pre-defined amount of \textit{gas}~\cite{Wo14}. When initiating a smart contract through sending a transaction, the sender has to specify the available $\textsf{gasLimit}$ that supports for execution, and the corresponding $\textsf{gasPrice}$ that the sender is willing to pay for each unit of gas. The transaction will get included in the blockchain successfully only when the balance of the sender is larger than $\textsf{gasLimit}\times\textsf{gasPrice}$. Although useful in avoiding network abuse, however, the gas system also sets some restrictions in designing our schemes as described in Section \ref{sec:scheme}.

\subsection{Smart Contract in Ethereum}\label{sec:smartcontract}

\textbf{Ethereum} is a new promising public blockchain platform~\cite{Wo14}. Its security is maintained by a cryptographic chain of puzzles (or blocks). Miners in the Ethereum network validate and approve transactions while mining new blocks. Mining a new block  by successfully solving a designated cryptographic puzzle rewards the miners with newly-created cryptocurrency and thus incentivizes them to mine more blocks, \ie, Proof-of-Work (PoW). The correctness of the network is guaranteed by this incentive mechanism. Anyone at any given point of time can join or leave/read/write/audit the public blockchain. In general, Ethereum provides us with two appealing properties:

\begin{itemize}
    \item \textit{Consensus.} The entire network agrees on the rules to verify each transaction and block. The data stored and computations executed on Ethereum must be consistent across miners and cannot be modified or denied.

    \item \textit{Transparency.} Ethereum is a public network. All the stored data and executed computations are transparent to any users.

\end{itemize}
Therefore, Ethereum acts as a trusted base who is \textit{trusted for correctness and availability, but not for privacy}.

\textbf{Smart contracts} in Ethereum are  applications with a state stored in the blockchain.  They can facilitate, verify, and enforce the process of a contract. Each smart contract, identified by a special address, consists of script code, a currency balance, and storage space in the form of a key/value store. Once created and deployed to Ethereum, the contract's code cannot be modified forever even for its creator.\footnote{Except for that a special \textit{suicide} opcode that clears all of the contract's data is used.} The contract can be triggered by a transaction from an external account or a call from other contracts, and is executed in transaction form. Once a smart contract transaction gets included in the blockchain, all the nodes in the network are expected to verify its validity by repeating the contract script. The most distinguished feature of smart contract in Ethereum lies in its support for \textit{Turing-complete scripting}, which makes it feasible for us to design various complex functions.

\subsection{Smart Contract in Hyperledger}\label{sec:smartcontractHyper}

\textbf{Hyperledger}  is a modular and extensible open-source system for deploying and operating private (or consortium) blockchains~\cite{Hyperledger}. It is a typical kind of permissioned blockchain, running among a set of known and identified participants who share a common goal but do not fully trust each other. Usually the consensus is guaranteed using traditional protocols like PBFT~\cite{castro1999practical}.

\textbf{Smart contracts} in Hyperledger, also called chaincodes, are supported to implement the application logic written
in general-purpose programming languages (\eg, Go, Java, Node.js). The execution of smart contracts in Hyperledger is different from that in Ethereum. In Hyperledger, instead of following the order-execute architecture, a new execute-order-validate architecture is realized to improve system efficiency and stability. Such design enables us to deploy a more efficient application. More importantly, no cryptocurrency is needed to support the execution of smart contracts.

\section{Security Definitions}

In this section, we explicitly discuss the security goals our design aims to achieve.

%Fairness has been extensively studied in secure multi-party computations~\cite{As14,GoIs10,GoHa11}. Ensuring fairness is also one of important security properties that we should achieve. Unfortunately, to date only restricted classes of functions are shown to be able to achieve completely fair secure computation~\cite{As14,GoHa11}, while partially fair secure computation for all functions can be constructed as shown in~\cite{BeLi11,GoKa10}. Inspired by~\cite{BeKu14}, we define fair SSE ($\mathcal{F}$-SSE) as follows.

\textbf{Soundness.} This property is derived from~\cite{BoFo16}
%and used to evaluate security in case of malicious adversary. Basically, it
which basically indicates that
the server will get caught if it tries to deviate from the protocol. In other words, the data owner (and other users) will not accept a wrong search result. Usually existing works achieve this objective by letting the data owner conduct a series of verifications.
%Basically, soundness is used to evaluate security in the case of malicious adversary where the data owner needs to verify the received search results.
In this paper, we extend this notion to claim that the received search results are reliable and correct definitely, and no verification is needed on the data owner.

\textbf{Confidentiality.} The confidentiality of SSE evaluates the private information protected from the adversary. It follows the real/ideal simulation paradigm~\cite{KaPa12,CuGa11,CaJa14} and is parametrized by three \textit{leakage functions} $\mathcal{L}=(\mathcal{L}_1,\mathcal{L}_2,\mathcal{L}_3)$ that describe what is allowed to leak to the adversary and are formalized as stateful algorithms. Formally, we have,

\begin{definition}
Let $\Pi=(\mathsf{Setup}$,$\mathsf{Search}$,$\mathsf{Update})$ be a dynamic SSE scheme and consider the following experiments with a stateful adversary $\mathcal{A}$, a stateful simulator $\mathcal{S}$ and three stateful leakage functions $\mathcal{L}=(\mathcal{L}_1$, $\mathcal{L}_2$, $\mathcal{L}_3)$:
\begin{description}
\item$\mathbf{Real}^{\Pi}_{\mathcal{A}}(\lambda):$  $\mathcal{A}$ chooses $\mathsf{DB}$. The challenger runs $\mathsf{Setup}(\textsf{DB})$ to generate the key $K$ and gives $\mathsf{EDB}$ to $\mathcal{A}$. Then $\mathcal{A}$ repeatedly makes $\mathsf{Search}$ and $\mathsf{Update}$ queries where $\mathcal{A}$ chooses challenger's input $\mathsf{in}$. Meanwhile, the experiment runs $\mathsf{Search}$ or $\mathsf{Update}$ with challenger's input $(K,\mathsf{in})$ and $\mathcal{A}$'s input $\mathsf{EDB}$, and gives the transcript to $\mathcal{A}$. Finally, $\mathcal{A}$ returns a bit $b$ as the output of the experiment.

\item$\mathbf{Ideal}^{\Pi}_{\mathcal{A}, \mathcal{S}}(\lambda):$  $\mathcal{A}$ chooses $\mathsf{DB}$. The simulator is given $\mathcal{L}_1(\mathsf{DB})$ and sends $\mathsf{EDB}$ to $\mathcal{A}$.
 Then $\mathcal{A}$ repeatedly makes $\mathsf{Search}$ and $\mathsf{Update}$ queries where $\mathcal{A}$ chooses simulator's input $\mathsf{in}$. Meanwhile, the experiment runs $\mathsf{Search}$  (resp. $\mathsf{Update}$) with simulator input $\mathcal{L}_2(\mathsf{in})$ (resp. $\mathcal{L}_3(\mathsf{in})$)   and gives the simulated transcript to $\mathcal{A}$.
Finally, $\mathcal{A}$ returns a bit $b$ as the output of the experiment.
%\end{itemize}

\end{description}

We say that $\Pi$ is \textit{$\mathcal{L}$-secure against adaptive attacks} if for all probabilistic polynomial-time (PPT) adversaries $\mathcal{A}$, there exists a probabilistic polynomial-time simulator $\mathcal{S}$ such that
\begin{equation*}
|\Pr[\mathbf{Real}^{\Pi}_{\mathcal{A}}(\lambda) = 1]-\Pr[\mathbf{Ideal}^{\Pi}_{\mathcal{A}, \mathcal{S}}(\lambda) = 1]| \leq \textsf{negl}(\lambda).
\end{equation*}

\end{definition}

The \textit{$\mathcal{L}$-secure against non-adaptive attacks} can be defined in the same way, except that in both experiments $\mathcal{A}$ must choose all of its queries at the start, and $\mathcal{L}$ takes them all as input and gives the output to $\mathcal{S}$ who generates \textsf{EDB} and the transcripts at the same time.

\section{Decentralized SSE in Public Blockchain}\label{sec:scheme}

We first construct a decentralized SSE  scheme $\Pi$ with off-the-shelf public blockchains. To give an exemplary instantiation, $\Pi$ is adapted from existing pioneering inverted index frameworks (such as~\cite{CaJa14,Ra16}) and modified to fit the decentralized environment. Therefore, soundness is automatically implied as long as the security of the underlying decentralized platforms is guaranteed. In Section~\ref{sec:general}, we show that the other SSE schemes with expressive queries or complex data types can also be extended to our settings similarly.

\subsection{Design Challenges and Countermeasures}

Intuitively, any traditional SSE scheme can be directly adapted to decentralized environment by replacing the central server with the smart contract. Unfortunately, some innovative features that guarantee the robustness and security of smart contract become obstacles instead in this adaption. Next we present some main design challenges and summarize the countermeasures at a high level.

\textit{Gas Limitation.} In Ethereum, each transaction that calls a function of the smart contract has a upper bound of consumed gas, called $\textsf{gasLimit}$ as described in Section~\ref{sec:smartcontract}. Each operation, including sending/storing data and executing computations, has a fixed gas cost. This restricts the designed function to have extremely limited computation steps and storage. Therefore, to make SSE over a large database become feasible, we are motivated to divide the database into smaller ones and conquer them individually. Simply speaking, in the setup phase where a large encrypted index is built, we partition the encrypted index into several blocks and upload them to the contract with sufficient transactions such that each transaction consumes less gas than $\textsf{gasLimit}$.  To ensure correctness, the contract needs to align the data together in order to return all matched results.
%Besides, to guarantee confidentiality, the built index will be randomly permutated before sent to the contract.

\textit{Gas Availability.} In the smart contract, each transaction is also associated with a $\textsf{gasPrice}$ that specifies the money the sender is willing to spend to purchase the gas.  It is required that the user who initiates the transaction has an account balance larger than the gas cost for executing the transaction. Otherwise the transaction will abort intermediately while the consumed gas cannot be refunded. Thus we should be very careful with the contract design with regard to gas cost. Particularly, it is critical to ensure that each functionality (\eg, \textsf{Search},\textsf{Update}) in the contract incurs lower gas cost than the sender's account balance.

\textit{The Verifier's Dilemma.} In Ethereum, miners are required to check the validity of transactions. However, verifying transactions may become significantly expensive when there are abundant and complex expressions in smart contracts. For rational miners, they are thus incentivized to skip the verification of the expensive transactions so as to stay ahead in the race to mine the next block. This phenomenon is called the \textit{verifier's dilemma}~\cite{LuTe15}. To mitigate this attack, we are motivated to reduce the computation burden on the contract as much as possible.
Our first observation is that the smart contract supports dictionary data type, and the main computation overhead of SSE lies in the search phase. In light of this, we make use of a dictionary to store encrypted index (\ie, \textsf{EDB}), which makes the search time complexity be $\mathcal{O}(d_w)$, where $d_w$ is the number of times that the keyword $w$ has been historically added to the database.
Our second optimization is the ultilization of packing method inspired by~\cite{CaJa14}. Specifically, we can pack multiple plaintexts and encrypt the output to obtain one ciphertext with the same size. The search result is thus in blocks instead of individuals. Besides, packing also helps us circumvent the above \textit{Gas Limitation} since it greatly reduces the storage cost. We note that although~\cite{CaJa14} claimed to use the packing method as well, it didn't describe how to implement it explicitly.

\begin{figure}[!t]
\centering
\includegraphics[width=0.7\columnwidth]{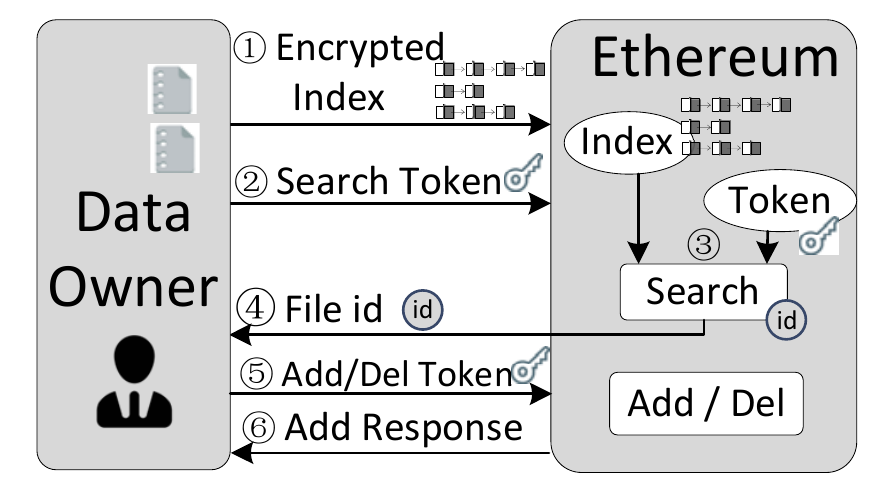}
%\vspace{-2mm}
\caption{A system overview for our scheme $\Pi$.}
\label{fig:system_overview}
\vspace{-2mm}
\end{figure}

\subsection{System Overview}

In Fig.~\ref{fig:system_overview}, we outline the architecture of our design. Then the data owner builds an encrypted index of keyword/identifier pairs and sends it to Ethereum, where complex computations are available via the smart contract.
For ease of presentation, operations on the data documents are not shown in the framework since the data owner could easily employ the traditional symmetric key cryptography to encrypt documents and then outsource encrypted data to any decentralized file storage network like InterPlanetary File System (IPFS).
The reason why we don't put encrypted documents on Ethereum is that it is very expensive to store data on it. Offloading huge data sets to another platform while
focusing on computation on Ethereum with small data storage benefits the Ethereum network greatly with regard to efficiency and robustness.

For each query, the data owner sends a transaction containing the search token to the designated smart contract. Note that each contract has a unique address in Ethereum. With the search token and previously stored index, the smart contract executes search algorithms and saves the search results (\ie, file identifiers) to its state. The data owner can later read the state and use file id to retrieve real documents from file storage network. For adding or deleting files, the data owner also only needs to send add/delete tokens to the contract and wait for the transactions to be mined into the block. For the add operation, our scheme requires the data owner to maintain a dictionary locally. In fact this is unnecessary and we could modify the scheme slightly to make the data owner stateless as shown in Section~\ref{sec:stateless}.

\begin{figure}[!t]

\fbox{
\begin{minipage} [t]{0.94\linewidth}
%\normalsize
%\small

%\vspace{5pt}
\noindent Our scheme $\Pi$: Protocol on the data owner.
\rule{\textwidth}{0.2mm}

\underline{\textbf{Setup}(\textsf{DB}):}
\vspace{-2pt}
\begin{enumerate}
\item Initialize an empty list $ \textsf{L}$, and an empty dictionary $\sigma$; Sample three keys $K, K^A, K^D\xleftarrow{\$} \{0,1\}^{\lambda}$.

%\item Sample three keys $K, K^A, K^D\xleftarrow{\$} \{0,1\}^{\lambda}$.% used for search, add and delete.

\item \textbf{For} each keyword $w \in \textsf{W}$:

\begin{enumerate}
\item $K_1 \leftarrow F(K,1||w)$;  $K_2 \leftarrow F(K,2||w)$;
\item Set $\alpha \leftarrow \lfloor \frac{|\textsf{DB}(w)|}{p}\rfloor, c \leftarrow 0$, where $p$ denotes the number of identifiers that can be packed.

\item Divide $\textsf{DB}(w)$ into $\alpha+1$ blocks. Pad the last block to $p$ entries if needed.

\item \textbf{For} each block in $\textsf{DB}(w)$:
\begin{itemize}[leftmargin = *]
  \item[-]  $\widetilde{\textsf{id}} \leftarrow \textsf{id}_1 || \textsf{id}_2 || ... || \textsf{id}_{p}$; $r \xleftarrow{\$} \{0,1\}^{\lambda}$.
  \item[-]  $d \leftarrow  \widetilde{\textsf{id}}\oplus G_{K_2}(r)$;  $l \leftarrow F(K_1, c)$; $c++$.
%  \item[-]  $l \leftarrow F(K_1, c)$.
  \item[-] Add $(l,d,r)$ to the list $\textsf{L}$ in lex order.% lexicographic order.
\end{itemize}

\end{enumerate}

%\item Set $\textsf{EDB} = \textsf{L}. $
\item Set $\textsf{EDB} = \textsf{L} $; Partition $\textsf{EDB}$ into $n$ blocks $\textsf{EDB}_i$ for $1 \leq i \leq n$, and send them to the contract. %  iteratively.

\end{enumerate}
%at \textsc{Client}
%\vspace{5pt}
%\textsc{Client}:

%
\rule{\textwidth}{0.2mm}

\underline{$\textbf{Search}(K,K^A,K^D,w)$}:

\vspace{-2pt}
\begin{enumerate}

%\item $K^A \leftarrow F(K,3)$, $ K^D \leftarrow F(K,4).$
\item  $K_1 \leftarrow F(K,1||w)$, $K_2 \leftarrow F(K, 2||w).$
\item $K^A_1 \leftarrow F(K^A,1||w)$, $K^A_2 \leftarrow F(K^A, 2||w).$
\item $K^D_1 \leftarrow F(K^D, w)$; $c \leftarrow 0$; Estimate $R$ and $\textsf{step}$.
\item \textbf{For} $i=0$ \textbf{to} $R$:
  \begin{enumerate}
    \item Send search token $ST=(K_1,K_2,K^A_1,K^A_2,K^D_1,c)$ to the contract.
    \item $c \leftarrow c+\textsf{step}$.
  \end{enumerate}
\end{enumerate}
%
%\vspace{5pt}

\rule{\textwidth}{0.2mm}

\underline{$\textbf{Add}(K,K^A,K^D,\textsf{id},\textsf{W}_{\textsf{id}})$} \textit{phase 1}:

\vspace{-2pt}
\begin{enumerate}

\item Initialize an empty list $ \textsf{L}^A$.

%\item $K^A \leftarrow F(K,3)$, $ K^D \leftarrow F(K,4).$

\item \textbf{For} each keyword $w \in \textsf{W}_{\textsf{id}}$:
\begin{enumerate}

\item  $K_1 \leftarrow F(K,1||w)$, $K_2 \leftarrow F(K, 2||w).$
\item $K^A_1 \leftarrow F(K^A,1||w)$, $K^A_2 \leftarrow F(K^A, 2||w).$
\item $K^D_1 \leftarrow F(K^D, w)$; $r \xleftarrow{\$} \{0,1\}^{\lambda}$.
\item $c \leftarrow \textsf{Get}(\sigma,w)$; If $c = \perp $ then $c \leftarrow 0$.
\item $l \leftarrow F(K^A_1,c)$; $d \leftarrow \textsf{id}\oplus G_{K^A_2}(r)$.
\item $\textsf{id}_{\textsf{del}} \leftarrow F(K^D_1,\textsf{id})$.
\item Add $(l,d,r,\textsf{id}_{\textsf{del}})$ to $ \textsf{L}^A$  in lex order. %lexicographic order.

\end{enumerate}

\item Send $ \textsf{L}^A$ to the contract.
\end{enumerate}

\underline{$\textbf{Add}(K,K^A,K^D,\textsf{id},\textsf{W}_{\textsf{id}})$} \textit{phase 2}:

\vspace{-2pt}
\begin{enumerate}
\item Read \textsf{re} from the contract.
\item \textbf{For} $i=0$ \textbf{to} $|\textsf{re}|$:
\begin{enumerate}
  \item \textbf{if} $\textsf{re}[i] = 0$:

  \begin{itemize}[leftmargin = *]
   \item[-] Fetch the $i$-th keyword $w$ in $\textsf{W}_{\textsf{id}}$.
   \item[-] $c \leftarrow \textsf{Get}(\sigma, w)$; c++; Insert $(w,c)$ into $\sigma$.
 %   \item[-] Insert $(w,c)$ into $\sigma$.
  \end{itemize}

\end{enumerate}
\end{enumerate}

\rule{\textwidth}{0.2mm}

\underline{$\textbf{Delete}(K^D,\textsf{id},\textsf{W}_{\textsf{id}})$}:

\vspace{-2pt}
\begin{enumerate}
\item Initialize an empty list $ \textsf{L}^D$.
\item \textbf{For} each keyword $w \in \textsf{W}_{\textsf{id}}$:

\begin{enumerate}
%\item $K^D \leftarrow F(K, 4)$;
\item $K^D_1 \leftarrow F(K^D,w)$, $\textsf{id}_{\textsf{del}} \leftarrow F(K^D_1, \textsf{id})$.
\item Add $\textsf{id}_{\textsf{del}}$ to $\textit{L}^D$  in lex order.

\end{enumerate}

\item Send $\textit{L}^D$ to the contract.
\end{enumerate}
\end{minipage}

}

%\end{tabular}
\caption{Our decentralized SSE scheme in Ethereum.}
\label{fig:schemeDO}
\vspace{-10pt}
\end{figure}

\begin{figure}[t]

\fbox{
\begin{minipage} [t]{0.94\linewidth}

\noindent Our scheme $\Pi$:  Protocol on the smart contract.

\rule{\textwidth}{0.2mm}
\underline{\textbf{Setup}($\textsf{EDB}_1, \textsf{EDB}_2, \cdots, \textsf{EDB}_n$):}
\vspace{-2pt}
\begin{enumerate}
\item Initialize two empty dictionaries $\gamma$ and $\gamma^A$.
\item Initialize an empty list $\textsf{ID}_{\textsf{del}}$.
\item \textbf{For} each received  $\textsf{EDB}_i$:
\begin{enumerate}
  \item Parse each entry in $\textsf{EDB}_i$ into $(l, d,r)$.
  \item Add each $(l, d||r)$ to $\gamma$.
\end{enumerate}

\end{enumerate}

\rule{\textwidth}{0.2mm}
\underline{$\textbf{Search}(K_1, K_2, K_1^A, K_2^A, K_1^D, c$}):
\vspace{-2pt}
\begin{enumerate}
\item Assert the estimated gas cost is lower than the balance.
\item \textbf{For} $i=0$ until \textsf{Get} returns $\perp$ or $i \geq \textsf{step}$:
\begin{enumerate}
\item  $l \leftarrow F(K_1, c)$; $d,r \leftarrow \textsf{Get}(\gamma, l)$.

\item  $\widetilde{\textsf{id}} \leftarrow d\oplus G_{K_2}(r)$; $c++$; $i++$.
\item Parse $\widetilde{\textsf{id}}$ into $(\textsf{id}_1, \cdots, \textsf{id}_{p})$.
\item Assert $\textsf{id}_j \notin \textsf{ID}_{\textsf{del}}$ ($1 \leq j \leq p$).
%\item Remove $\textsf{id}_j$.
\item Save $\textsf{id}_j$ to the state.
\end{enumerate}

\item Assert $\gamma^A$ has not been searched.
\item \textbf{For} $c=0$ until \textsf{Get} returns $\perp$:
\begin{enumerate}
\item  $l \leftarrow F(K_1^A, c)$; $d,r \leftarrow$ Get$(\gamma^A, l)$.
\item  $\textsf{id} \leftarrow d\oplus G_{K^A_2}(r)$; $c++$.
\item Assert $\textsf{id} \notin \textsf{ID}_{\textsf{del}}$.
\item Save $\textsf{id}$ to the state.
\end{enumerate}
\end{enumerate}

\rule{\textwidth}{0.2mm}
\underline{$\textbf{Add}(\textsf{L}^A$}):
\vspace{-2pt}
\begin{enumerate}
\item Initialize an empty list \textsf{re} of size $|\textsf{L}^A|$.
\item Parse each tuple of $\textsf{L}^A$ into $(l, d, r, \textsf{id}_{\textsf{del}}$).
\item Set $i\leftarrow 0$.
\item \textbf{For} each tuple in $\textsf{L}^A$:
\begin{enumerate}
\item \textbf{if} $\textsf{id}_{\textsf{del}}\in \textsf{ID}_{\textsf{del}}$:
\begin{itemize}[leftmargin = *]
\item[-]  $\textsf{re}[i] \leftarrow 1$.
\item[-] Delete $\textsf{id}_{\textsf{del}}$ from $\textsf{ID}_{\textsf{del}}$.
\end{itemize}
\item \textbf{else}:
\begin{itemize}[leftmargin = *]
\item[-] $\textsf{re}[i] \leftarrow 0$.
\item[-] Add $(l, d||r)$ to $\gamma^A$.
\end{itemize}
\item $i++$.
\end{enumerate}
\item Save \textsf{re} to the state.

\end{enumerate}

\rule{\textwidth}{0.2mm}
\underline{$\textbf{Delete}(\textsf{L}^D)$}:
\vspace{-2pt}
\begin{enumerate}
\item \textbf{For} each element $\textsf{id}_{\textsf{del}}$ in $\textsf{L}^D$:
\begin{enumerate}
  \item Add $\textsf{id}_{\textsf{del}}$ to $\textsf{ID}_{\textsf{del}}$.
\end{enumerate}
\end{enumerate}

\end{minipage}
}
\caption{Our decentralized SSE scheme in Ethereum.}
\label{fig:basicSmart}
\vspace{-10pt}
\end{figure}

\subsection{Our Detailed Construction}

In Fig.~\ref{fig:schemeDO} and Fig.~\ref{fig:basicSmart}, we give a formal description of our decentralized SSE scheme $\Pi$.
%$\Pi$ is secure against non-adaptive attacks.
For simplicity, let $F: \{0,1\}^{\lambda}\times\{0,1\}^*\rightarrow\{0,1\}^{\lambda}$, $G: \{0,1\}^{\lambda}\times\{0,1\}^{\lambda}\rightarrow\{0,1\}^{*}$ be two pseudo-random functions (Note that there should be different PRFs for different input keys).
%Let (\textsf{Enc}, \textsf{Dec}) be a symmetric-key encryption scheme that is secure against chosen-plaintext attack (CPA).
We use $||$ to denote the concatenation operation.  ``$\lfloor \cdot \rfloor$'' is a floor function, and ``$|\cdot|$'' denotes the number of elements in a list. For a dictionary data type, it includes two algorithms: \textsf{Add} and \textsf{Delete}. And we use term \textsf{Get} to fetch the specified data item in a dictionary. For example, given a dictionary data type $\gamma$ and an input label $l$, $\textsf{Get}(\gamma, l)$ outputs the corresponding item $d||r$ and parses it into $d$ and $r$.

In the \textsf{Setup} phase, the data owner divides $\textsf{DB}(w)$ into $\alpha+1$ blocks, with each block of $p$ entries. Here $p$ is a system parameter chosen by the data owner. We use concatenation to pack multiple file identifiers into one. To ensure confidentiality, the bit length of $\widetilde{\textsf{id}}$ should be less than that of the security parameter $\lambda $. Therefore, we have $p \leq \frac{\lambda }{l}$, where $l$ is the bit length of the file identifier. Note that before uploading the database, the list $\textsf{L}$ should be placed in lexicographic order. Otherwise it will leak information about the order in which the input was processed. To avoid exceeding $\textsf{gasLimit}$, we partition the encrypted database into $n$ blocks and send them to the contract one by one with $n$ different transactions. At the contract side, they are received iteratively and placed together using dictionary data type. Similarly, the search process will be completed with $R$ transactions, each of which returns $\textsf{step}$ items at most. Here $n$, $R$ and $\textsf{step}$ are public system parameters and experimentally determined.

In the \textsf{Add} phase, we encrypt file id without using packing. This is because encrypting several plaintexts into one ciphertext makes it hard for the contract to identify which file/keyword pair has been previously deleted, \ie, whether it exists in the set $\textsf{ID}_{\textsf{del}}$. In addition, in reality changes often happen with only one or several documents at one time. Update incurs much less gas cost than the \textit{Gas Limitation.} Therefore, individually dealing with file id satisfies the system requirements for update operations.

For the protocol on the smart contract, we remark that transaction triggering functions in smart contract doesn't return any results. Execution of any function only changes its state that is permanently stored on Ethereum. We implement our scheme by saving search results into the state and later reading them on the data owner side.

\subsection{Multi-user Setting}

In this work, we further address the issue of multi-user data sharing as considered in~\cite{JaJu13,CuGa11,KiOk16}. In such applications, the data owner is interested in allowing a third party (\ie, other users) to search the database, while the other users learn the information that the data owner authorizes them to learn but nothing else. The private information about the queries and search results should be protected from the adversary as well.

\begin{figure}[!t]
\centering
\includegraphics[width=0.7\columnwidth]{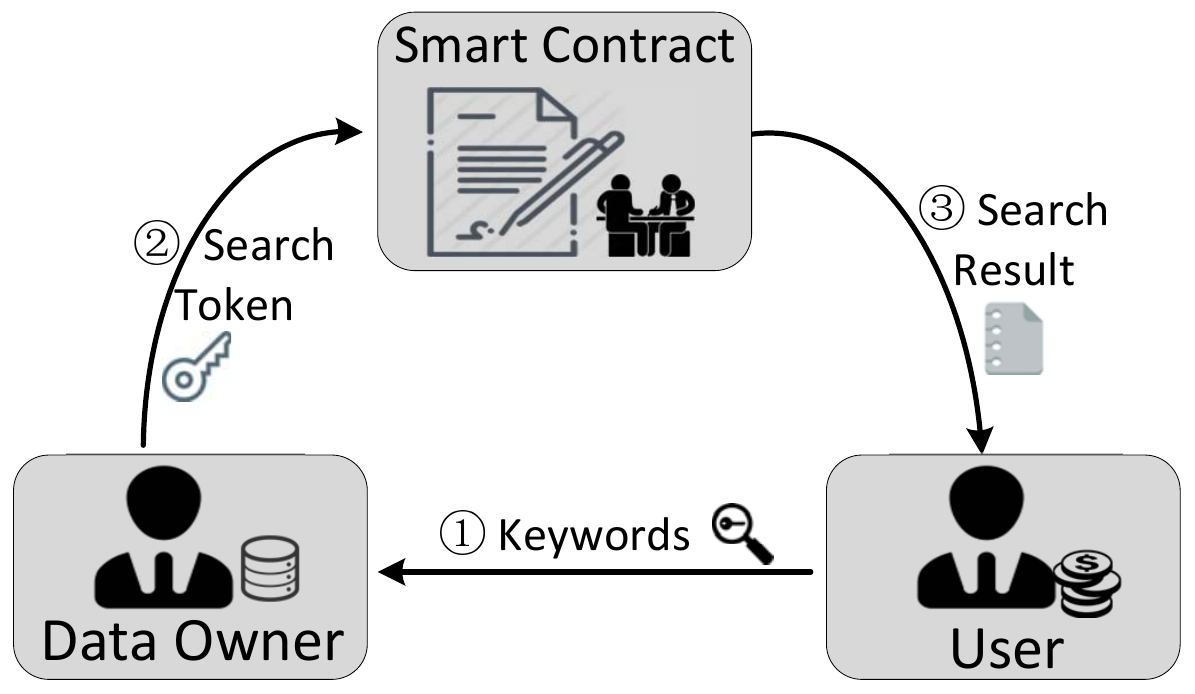}
%\vspace{-2mm}
\caption{System overview of multi-user setting in public blockchain.}
\label{fig:system_model}
\vspace{-2mm}
\end{figure}

Using existing cryptographic tools such as broadcast encryption~\cite{KiOk16}  is a possible solution to help the data owner add and revoke users.
In a permissionless blockchain environment like Ethereum, however, anyone at any time can participate in the network and read/write history records, and everything on the smart contract is public. It is not applicable to leverage such cryptographic schemes which usually require the nodes in the network to store a private key and perform decryption operations.
Currently we propose to use the straightforward extension for $\Pi$ as indicated in~\cite{JaJu13}: the data owner receives the user's query, and generates the corresponding search tokens as if himself is searching the database. Fig.~\ref{fig:system_model} gives an overview for the multi-user design. Relying on cryptographic tools  in a public blockchain environment to efficiently realize users searching and flexibly add/revoke users is a challenging problem and we leave it to our future work.

\section{Decentralized SSE in Private Blockchain}\label{sec:schemePri}

To expand the application scenarios, we construct $\Pi^+$ with the private blockchain, where a set of known and identified service providers (\ie, peer nodes) is available. Although bearing a stronger assumption for the blockchain network, $\Pi^+$ enjoys a higher efficiency than $\Pi$.
%Similarly, $\Pi^+$ is also constructed based on inverted index framework as did in $\Pi$.

\subsection{The Practical Concerns}

The private blockchain, such as Hyperledger, runs among a set of participants who do not trust each other but have a common goal and try to provide a service collaboratively. We emphasize that the assumption of such consortium holds in practice. Taking health  information sharing for example, a number of hospitals, research institutes, banks, and insurance companies may facilitate collaboration to maintain a shared medical database so as to provide a better user experience for patients. Typical examples include WorldCare~\cite{Worldcare}, OMAHA~\cite{omaha}, \etc
Building a private blockchain among these participants creates a transparent and reliable environment for medical data.
 Clinics or patient individuals can outsource their medical records, in encrypted versions, to the consortium for ease of management. When necessary, any participant from the consortium, after getting authorized by the data owner, can decrypt database locally and obtain correct medical information with assurance. In such application scenario, the participants in the consortium enjoy the benefit of a trusted database when getting access to. Search services with privacy preserved should also be supported by the consortium before the data owner releases private information to all the participants.

\subsection{Our Construction}
Although using different blockchain platforms with $\Pi$, we can regard blockchain as a black box and construct $\Pi^+$  similarly.
$\Pi^+$ is  constructed based on inverted index framework as well. The difference lies in the way we deal with the large data set.

In the Setup, $\Pi^+$ also divides $\textsf{EDB}$ into $n$ blocks. In Hyperledger fabric, however, there is a size limitation of the parameters. Generally speaking, we have $n=\frac{|\textsf{EDB}|}{\Delta_{\textsf{fabric}}}$, where $\Delta_{\textsf{fabric}}$ denotes the limitation for parameter size. According to our experiments, we can include as many as 500 entries of $\textsf{L}$ in one transaction, which is much more than that in $\Pi$.

In the Search step, since there is no gas limitation in private blockchain, we can query records as many as possible. Therefore $\Pi^+$ sets no limitation for $\textsf{step}$ and set $R=1$. In other words, the search token $ST$ is sent to the smart contract in one transaction, and the smart contract can execute search operation at a time.

$\Pi^+$ supports update operations over a large-scale data set. Similar to the construction in Fig.~\ref{fig:schemeDO}, $\Pi^+$ makes use of different secret keys to realize add or delete, \ie, using $K^A$ and $K^D$ to generate add token and delete token respectively. Besides, it is able to deal with large data set by using divide-and-conquer method, as did in the Setup phase. Our experiments will show that $\Pi^+$ supports adding several hundreds of files.

\begin{figure*}[!t]

\fbox{
\begin{minipage} [t]{\linewidth}
%\normalsize
%\small

%\vspace{5pt}
%\noindent Our basic scheme $\Pi$: Protocol on the data owner.
%
%\rule{\textwidth}{0.2mm}

\underline{\textbf{Setup}($n, \lambda$):}
\vspace{-2pt}
\begin{enumerate}
\item The data owner runs $(pk,\{sk_1,sk_2,\ldots,sk_n\}) \leftarrow \textsf{Init}(n, \lambda)$, and generates a secret key $\delta \leftarrow \{0,1\}^{\lambda}$.

%\item Sample three keys $K, K^A, K^D\xleftarrow{\$} \{0,1\}^{\lambda}$.% used for search, add and delete.

\item The data owner constructs a subset $S \subseteq \{1, 2, \ldots, n\}$ and performs broadcast encryption $\textit{Hdr} \leftarrow \textsf{Enc}_{S, pk}(\delta)$.

\item Sends $(\delta, \textit{Hdr})$ to the smart contract, and $(pk, sk_i)$ to user $i$.

\end{enumerate}

\rule{\textwidth}{0.2mm}

\underline{$\textbf{RevokeUser}(i, R, S)$}:

\vspace{-2pt}
\begin{enumerate}
\item The data owner generates a new secret key $\delta' \leftarrow \{0,1\}^{\lambda}$, and sets $S' \leftarrow S\backslash R\bigcup \{i\}$, $\textit{Hdr}' \leftarrow \textsf{Enc}_{S', pk}(\delta')$.

\item  Sends $(\delta', \textit{Hdr}')$ to the smart contract, which overwrites the old values of $(\delta, \textit{Hdr})$, respectively.

\end{enumerate}

\rule{\textwidth}{0.2mm}

\underline{$\textbf{AddUser}(i)$}:

\vspace{-2pt}
\begin{enumerate}

\item The data owner allocates an identity to user $i$, and sends $(pk, sk_i)$ to it.
\end{enumerate}

\rule{\textwidth}{0.2mm}

\underline{$\textbf{Trapdoor}(ST, i, pk, sk_i)$}:

\vspace{-2pt}
\begin{enumerate}
\item User $i$ downloads $S$ and $\textit{Hdr}$ from the blockchain, and computes $\widetilde{\delta} \leftarrow \textsf{Dec}_{S, i, pk, sk_i}(\textit{Hdr})$.

\item With search token $ST$, user $i$ computes $\widetilde{ST} \leftarrow ST \oplus \widetilde{\delta}$, and sends $\widetilde{ST}$ to the smart contract.
\end{enumerate}

\rule{\textwidth}{0.2mm}

\underline{$\textbf{Search}(\widetilde{ST}, \delta)$}:

\vspace{-2pt}
\begin{enumerate}
\item The smart contract computes $St \leftarrow \widetilde{ST} \oplus \delta$. If $ST$ is a valid search token, then executes $\textbf{Search}(ST)$ and outputs the result, otherwise returns $\perp$.

\end{enumerate}

\end{minipage}

}

%\end{tabular}
\caption{Multi-user SSE construction in private blockchain.}
\label{fig:multiuser}
\vspace{5pt}
\end{figure*}

\subsection{Multi-user Setting}
Different from public blockchain environment, only authenticated participants are allowed to join in the private blockchain network.
In light of this, we propose making use of broadcast encryption~\cite{boneh2005collusion} to facilitate multi-user data sharing for $\Pi^+$. A broadcast encryption system consists of three randomized algorithms $(\textsf{Init}, \textsf{Enc}, \textsf{Dec})$. $\textsf{Init}(n)$ takes as input the number of users and outputs a public key $pk$ and $n$ secret keys $\{sk_1, sk_2, \ldots, sk_n \}$. $\textsf{Enc}$ takes as input a subset $S$ and a public key, and outputs the broadcast ciphertext $\textit{Hdr}$. $\textsf{Dec}$ takes as input a subset $S$, a user id $i \in \{1, 2, \ldots, n\}$, public key $pk$, the private key $sk_i$ for user $i$, and  a broadcast ciphertext. It outputs the plaintext if $i \in S$. Our multi-user construction is illustrated in Fig.~\ref{fig:multiuser}. Compared with the single-user scheme, the contract only needs to perform some extra simple operations (\ie, xor) in order to determine if the user has been revoked. It is very efficient in practice.
Our multi-user design requires that the peers executing smart contracts maintain a private key $\delta$. Such requirement is easy to realize since every participant in the private blockchain is identified and permitted by others to join the consortium. They are motivated to maintain their reputation and not likely to take the risk of colluding with users and revealing the secret key.

\begin{table*}[!t]
%\normalsize

\begin{center}
\begin{tabular}{c|c|c|c|c|c|c|c}
\hline
Scheme & Consensus Algorithm & Mining  & Scalability & Efficiency  & Performance Bottleneck & Privacy Guarantee & Trustworthy\\
\hline
$\Pi$ & PoW & Yes & High & Low & Mining Process & Yes & High\\
\hline
$\Pi^+$ & PBFT & No & Low & High &  Database Size & Yes & Low \\
\hline
\end{tabular}
\end{center}
\caption{A comparison between our two designs}
\label{tab:comparison_two_solution}
\vspace{-10pt}
\end{table*}

\section{Theoretical Analysis}

\subsection{Comparison Between Our Two Designs}

Our two proposed decentralized schemes make use of two different kinds of blockchains, leading to a trade-off between security and efficiency.
$\Pi$ is constructed over public blockchain  which already provides a decentralized computing platform.
It enjoys a high scalability since everyone can freely access to the public platform and build their own SSE applications.
However, its consensus is guaranteed through costly PoW-based mining process, which becomes the main performance bottleneck for $\Pi$. Specifically, for each transaction that triggers search or update function, only after the transaction gets included into the valid block should we have confidence in the correctness of search results. Currently it takes about $17s$  to mine a block in Ethereum, which means that we have to wait $17s$ until we could get the search results. A detailed explanation is presented in Section~\ref{sec:exper}.

On the other hand, $\Pi^+$ requires a stronger security assumption of a consortium, which has limited application scenarios. Unlike public blockchain that trusts the whole world, $\Pi^+$ believes that the entire consortium is trusted and always generates correct data.
Due to the high efficiency of private blockchain resulting from the fast consensus algorithm (\eg, PBFT), its performance is mainly affected by the database size, as shown in our experiments in Section~\ref{sec:expSim}. The time complexity of $\Pi^+$ is $\mathcal{O}(d_w)$ for a search and $\mathcal{O}(|\textsf{W}_{\textsf{id}}|)$ for an update.
Table~\ref{tab:comparison_two_solution} presents a concise comparison between them. We emphasize that $\Pi$ has a higher trustworthy degree than $\Pi^+$ since the public blockchain relies on the assumption that the majority of the whole world are honest, while the private blockchain assumes the majority of the involved participants to be honest.
We believe that corrupting more users (\ie, 50\% of the whole world vs. 50\% of a set of participants) is much more difficult, since it needs to unite more network nodes for the collusion purpose.

\subsection{Security Analysis}

\textbf{Soundness:} It is straightforward to see that soundness can be achieved as long as the security of blockchain is guaranteed. This is because if smart contracts are correctly executed on blockchains, the search results will be stored as contract states permanently. Each node in the blockchain network can verify the states. The \textit{consensus} property of blockchain ensures the correct execution of each search operation.

\textbf{Confidentiality:}
Since $\Pi$ and $\Pi^+$ have similar system model and desgin goal (\ie, protecting database from adversary), we will only present a security proof sketch for $\Pi$ and the security of $\Pi^+$ can be proved similarly.
 To prove confidentiality, we first proceed with the formal definition of three stateful leakage functions $\mathcal{L}=(\mathcal{L}_1$, $\mathcal{L}_2$, $\mathcal{L}_3)$ considered in our construction. Amongst the state, a list $Q$ recording all queries that have been submitted will be maintained. Specifically, each entry of the list $Q$ is of the form $(i,\mathsf{op},\ldots)$, where $i$ denotes a counter, $\mathsf{op}$ denotes the operation type, and the rest denote the inputs to the operation.
\begin{itemize}
\item (\textit{Leakage function $\mathcal{L}_1$}). Given an initial input $\mathsf{DB}$, $\mathcal{L}_1(\mathsf{DB})=\sum_{w\in \mathsf{W}}\lceil \frac{|\mathsf{DB}(w)|}{p} \rceil$. Meanwhile, it initializes a counter $i=0$, an empty list $Q$, a set $\mathsf{ID}$ containing all the identifiers in $\mathsf{DB}$, and saves them as the state.
\item (\textit{Leakage function $\mathcal{L}_2$}). Given a search input $w$, $\mathcal{L}_2(\mathsf{in})=\{ \mathsf{sp}(w,Q), \mathsf{DB}(w), \mathsf{AP}(w,Q,\mathsf{ID}), \mathsf{DP}(w,Q,\mathsf{ID}) \}$, where $\mathsf{sp}(w,Q)$ denotes the search pattern, $\mathsf{AP}(w,Q,\mathsf{ID})$ (resp. $\mathsf{DP}(w,Q,\mathsf{ID})$) denotes the add (resp. deletion) pattern of the keyword $w$ with respect to $Q$ and $\mathsf{ID}$, all of which are defined below. Meanwhile, it increases $i$ and appends $(i,\mathsf{search},w)$ to $Q$.
\item (\textit{Leakage function $\mathcal{L}_3$}). Given an add update input $(\mathsf{id},\mathsf{W}_{\mathsf{id}})$, $\mathcal{L}_3(\mathsf{in})=\{ \mathsf{add}, |\mathsf{W}_{\mathsf{id}}|, (\mathsf{sp}(w,Q),\mathsf{ap}(\mathsf{id},w,Q), \\ \mathsf{dp}(\mathsf{id},w,Q)):w \in \mathsf{W}_{\mathsf{id}}\}$, where $\mathsf{ap}(\mathsf{id},w,Q)$ (resp. $\mathsf{dp}(\mathsf{id},w,Q)$) denotes the add (resp. deletion) pattern of $\mathsf{id},w$ with respect to $Q$, both of which are defined below. Meanwhile, it increases $i$, appends $(i,\mathsf{add},\mathsf{id},\mathsf{W}_{\mathsf{id}})$ to $Q$ and adds $\mathsf{id}$ to $\mathsf{ID}$. For a delete update input, the only difference is that $\mathcal{L}_3(\mathsf{in})$ outputs $\mathsf{del}$ instead of $\mathsf{add}$ as the first component. Finally, if any of the search patterns was non-empty, then it also outputs $\mathsf{id}$.
\end{itemize}

Here, we define all the patterns mentioned above. The search pattern $\mathsf{sp}(w,Q)$ is a set of indices of queries where $w$ was searched for, \ie, $\mathsf{sp}(w,Q)=\{j:(j,\mathsf{srch},w)\in Q\}$. Namely, the search pattern reveals whether the keyword $w$ has been searched before. The add pattern $\mathsf{ap}(\mathsf{id},w,Q)$ is the set of indices where $w$ was added to the document $\mathsf{id}$, \ie, $\mathsf{ap}(\mathsf{id},w,Q)=\{j:(j,\mathsf{add},\mathsf{id},\mathsf{W}_{\mathsf{id}})\in Q,w\in \mathsf{W}_{\mathsf{id}}\}$. The add pattern $\mathsf{AP}(w,Q,\mathsf{ID})$ is the set of identifiers to which $w$ was added along with the indices showing when they were added, \ie, $\mathsf{AP}(w,Q,\mathsf{ID})=\{(\mathsf{id},\mathsf{ap}(\mathsf{id},w,Q)): \mathsf{id} \in \mathsf{ID}, \mathsf{ap}(\mathsf{id},w,Q) \neq \emptyset \}$. Besides, the deletion patterns $\mathsf{dp}(\mathsf{id},w,Q)$ and $\mathsf{DP}(w,Q,\mathsf{ID})$ can be defined analogously.

\begin{theorem}
If $G$ and $F$ are pseudo-random, then our scheme $\Pi$ is $\mathcal{L}$-secure against non-adaptive attacks.
\end{theorem}

Proof is deferred to Appendix~\ref{sec:append} for ease of exposition.

\section{Construction Variants}\label{sec:stateless}

\subsection{Adaptive Security}
$\Pi$ is proved to be secure against non-adaptive attacks. As is noted in~\cite{CaJa14}, making use of random oracle enables us to achieve adaptive security easily. Specifically, in $\Pi$ we replace the PRF $F$ with the random oracle $H$. For an input $m\in \{0,1\}^{\lambda}$ with key $K$, $F(K,m)$ is replaced with $H(K||m)$. And $G_K(r)$ is replaced with  $H(K||r)$ where $r$ is randomly chosen from $\{0,1\}^{\lambda}$. This variant has the same leakage function with $\Pi$. In the security proof, the simulator $\mathcal{S}$ also behaves similarly except that $\mathcal{S}$ needs to program the response of the random oracle in a way that it matches the query results that are already revealed. For the label $l$, $\mathcal{S}$ can set the response of $H(K||m)$ to be a random value with $\lambda$ bits in length. For the ciphertexts of $\textsf{id}$, $\mathcal{S}$ can set the random oracle such that the ciphertexts will be decrypted to the revealed results.

\subsection{Forward Privacy}
Forward privacy is also an important security design goal in SSE. It means that the adversary does not learn if the newly-added document contains a keyword that has been searched before. Inspired by recent progress~\cite{Ra16}, our designs can be easily extended to achieve forward privacy as well. The key idea is to use trapdoor permutation to make the search token unlinkable to the update token. Specifically, when generating a label for the $c$-th entry in $\textsf{DB}(w)$, instead of using a counter $c$ that increases itself, we use a trapdoor permutation $\pi$ in a way that $\beta_{c}=\pi^{-1}_{sk}(\beta_{c-1})$ and set the label as $l=F(K,\beta_c)$ where $\beta_0$ is a randomly chosen integer. Then on the smart contract, it can only compute $\beta_{c-1}=\pi_{pk}(\beta_{c})$ with the public key in polynomial time, but not $\beta_{c+1}$ since it has no secret key. Therefore, the $(c+1)$-th newly-added entry to $\textsf{DB}(w)$ without having been searched cannot be deduced from previously-leaked search token $\beta_c$. This variant has the same communication complexity with $\Pi$ (or $\Pi^+$), and the computation overheads on the data owner and the contract increase a little caused by permutation computation.

\subsection{Stateless Data Owner}
Currently our schemes require the data owner to maintain a local dictionary $\sigma$ consisting of  a counter for each keyword that is added after initialization. We could slightly modify the \textsf{Add} protocol to make the data owner stateless by encrypting $\sigma$ and sending the ciphertexts to any decentralized file storage systems (\eg, IPFS). The data owner can fetch the encrypted $\sigma$ and decrypt it for each $\textsf{Add}$ operation. The size of $\sigma$ relies on the number of distinct keywords that have been added in the \textsf{Add} phase, which is much smaller than the total number of keywords. In this case, the adversary can learn how many of new keywords were added into the database. This leaked information is acceptable in practice as far as we can see.

\subsection{Security Against Malicious Data Owner}

In the multi-user setting, $\Pi$ is vulnerable to a malicious data owner who arbitrarily reveals a random  search token. To mitigate such attack, we can use zero-knowledge proof~\cite{PaHo13} to force the data owner to reveal a correct search token. Specifically, we first let the data owner generate a proof for his search token by using zero-knowledge proof. Then we use smart contract to verify the proof, as did in~\cite{AhAn16}. If the search token is invalid we stop searching. In this way, the data owner earns nothing with the cheating.

\section{Generalization of our Framework}\label{sec:general}

In this work, we use smart contract to construct a decentralized SSE scheme based on the inverted index. We remark that many other SSE schemes
%that support complex queries or data structures
fit for our framework as well and can be extended to construct abundant decentralized SSE schemes with soundness guaranteed.

Recent works on SSE have focused on increasing their expressiveness such as supporting boolean queries~\cite{CaJa13,FaJa15,KaMo17}, or developing structured encryption like graph encryption~\cite{ChKa10,MeKa15}. All of them are also bothered with a serious security challenge: a malicious central server can output partial or even incorrect results whenever it wants. To address this concern,
these works can be tuned into our decentralized setting likewise.
The most intuitive observation of this extension is that smart contracts actually provide us with a trusted and transparent ``server''.
%, relying on which rich decentralized SSE schemes can be built.
The main obstacle lies in dealing with various limitations of gas system in smart contract when using public blockchain.
Our proposed several countermeasures (\eg, dividing the encrypted index and conquering them individually, packing multiple identifiers) throw light on how to address these issues.
Once constructed via smart contracts, the scheme is guaranteed with soundness  and thus there is no need to concern itself with a malicious server any more.

Storing data and executing computations in blockchain-based decentralized environments are  reliable and immutable.
We strongly believe that using decentralized platforms instead of a central server benefits a lot for the security requirements of SSE.

\begin{table}[t]
\centering
\begin{tabular}{|c|c|c|c|}
\hline
$\textbf{DB name}$ & $(\omega,\mathsf{id})$ $\textbf{pairs}$ & $\textbf{distinct keywords}$ & $\textbf{EDB}$\\
\hline\hline
\textsf{DB1} & $100,763$ & $22,673$& $5.4$MB \\
\hline
\textsf{DB2} & $300,617$ & $54,980$ & $14.1$MB\\
\hline
\textsf{DB3} & $500,567$ & $75,924$ & $21.3$MB\\
\hline
\textsf{DB4} & $1,000,141$ & $123,912$ & $39$MB\\
\hline
\end{tabular}
\caption{Evaluation database sizes.}
\label{tab:database}
%\vspace{-5pt}
\end{table}

\section{Implementation and Evalutations}\label{sec:exper}

We implement prototypes for both $\Pi$ and $\Pi^+$. We first evaluate $\Pi$ and $\Pi^+$ in local simulated blockchain networks with TestRPC and Hyperledger fabric, respectively.
Besides,  the multi-user design of $\Pi^+$ is evaluated as well to demonstrate the performance of adding/revoking users.
Considering the open property of public blockchain, we further deploy  $\Pi$ to an official Ethereum test network Rinkeby.

%In order  to focus on the performance analysis for our novel designs, we only evaluate $\Pi$ and $\Pi^+$ in the single-user setting and leave multi-user parts to our future work.

\begin{table*}[t!]
\begin{center}
\begin{tabular}{|c|c|c|c|c|c|c|c|c|c|}

\hline
\multirow{2}{*}{DB} & \multicolumn{3}{c|}{Setup} & \multicolumn{3}{c|}{Search} & \multicolumn{3}{c|}{Update} \\
\cline{2-10}
 & D.O. time & $\Pi$ & $\Pi^+$ & D.O. time & $\Pi$ & $\Pi^+$ & D.O. time & $\Pi$ & $\Pi^+$ \\
\hline
DB1 & 9s & 23min & 2min & $\approx 1ms$ & 7s & 0.3s & $\approx 1ms$ & 10s & 0.4s \\
\hline
DB2 & 15s & 66min & 6min & $\approx 1ms$ & 8s & 0.5s & $\approx 1ms$ & 10s & 0.4s \\
\hline
DB3 & 18s & 114min & 9min & $\approx 1ms$ & 10s & 0.6s & $\approx 1ms$ & 10s & 0.4s \\
\hline
DB4 & 23S & 949min & 15min & $\approx 1ms$ & 16s & 0.8s & $\approx 1ms$ & 10s & 0.4s \\
\hline
\end{tabular}
\end{center}
\caption{Evaluations for $\Pi$ and $\Pi^+$ in local simulated network. Here D.O. represents the time costs on the `Data Owner'. Search time is evaluated by returning 100 matched documents. Update overheads are given by adding and deleting a file, the size of which is chosen to incur only one transaction.}
\label{tab:timecost_rpc}
%\vspace{-2mm}
\end{table*}

\subsection{Implementation Details}

The data owner is instantiated on a local machine with 16GB of RAM, 4 Intel cores i7-3770, running Ubuntu 16.04.2.
For $\Pi$, we deploy the smart contract to a local simulated network TestRPC and also an official Ethereum test network Rinkeby, respectively. The data owner side is written in Python and the smart contract is implemented in Solidity in combination with Javascript as the intermediate interactive language. We implement PRF and random oracles using HMAC-SHA256. Since Ethereum currently does not support HMAC instantiation, we follow the standard construction of HMAC~\cite{KaLi14} and implement HMAC-SHA256 using Python and Solidity, respectively.
To avoid exceeding \textsf{gasLimit}, in the setup phase the encrypted database $\textsf{EDB}$ is divided into $n$ subsets and sent to the smart contract with $n$ transactions.
Due to the time-varying nature of $\textsf{gasLimit}$, we experimentally include 70 entries from the list $\textsf{L}$ in each transaction
%and thus we have $n=\frac{N}{70p}$ where $N$ is the total number of keyword/identifier pairs,
and set the pack number to be $p=8$. In addition, each search query is also completed with $R$ transactions at most, each of them returns $\textsf{step}=47$ items at most. In our experiments, $R=4$ satisfies our requirements.

For $\Pi^+$, we use the Hyperledger fabric framework  to construct a local private blockchain, and the smart contract (also named chaincode in Hyperledger) is written in Go language.
There are two peers in our test network belonging to different organizations and we use the default 256-bit ECDSA scheme for signature purpose.
We also use built-in HMAC-SHA256 function library supported by Go language. Due to the high scalability of private blockchain, we set $p=10$ and include 500 entries from the list $\textsf{L}$ in each transaction. One transaction is sufficient to complete search query and thus we set $R=1$ and no limitation for $\textsf{step}$.

The experiments reported in this work use datasets derived from Enron emails\footnote{https://www.cs.cmu.edu/\~./enron.}, which are a collection of plain text files. We extract a subset of emails and select increasing subsets from the original subset as document collections with different numbers of $(w,\textsf{id})$ (\ie, keyword/identifier) pairs. The key attributes of these datasets are summarised in Table~\ref{tab:database}.

\begin{table}[t]
    \caption{Efficiency evaluations for the multi-user setting}
    \begin{center}
        \begin{tabular}{c|c|c|c|c|c}
            \Xhline{1.2pt}
            % \toprule

            Number  & \multirow{2}*{\textbf{Setup}} & \multirow{2}*{\textbf{RevokeUser}} & \multirow{2}*{\textbf{AddUser}} & \multirow{2}*{\textbf{Trapdoor}} & \multirow{2}*{\textbf{Search}} \\
            of users & & & & & \\
            \hline
            8 & 0.21s & 0.19s & $\approx$ 1ms & 9.5ms & $\approx$ 1ms\\
            \hline
            64 & 1.06s & 0.92s &  $\approx$ 1ms & 9.6ms & $\approx$ 1ms\\
            \hline
            512 & 7.76s & 7.70s &  $\approx$ 1ms & 9.1ms &  $\approx$ 1ms\\
            \hline
            4096 & 63.32s & 63.26s & $\approx$ 1ms & 9.7ms & $\approx$ 1ms\\
            \hline
            32768 &  497.43s &496.29s& $\approx$ 1ms   & 9.3ms &  $\approx$ 1ms\\
            \Xhline{1.2pt}
        \end{tabular}
        \label{multiuser}
    \end{center}
\end{table}

\subsection{Experiments on Simulated Network}
\label{sec:expSim}
We first evaluate $\Pi$ and $\Pi^+$ on local simulated networks.
We use $\textit{TestRPC}$\footnote{https://github.com/ethereumjs/testrpc.}  to construct a simulated Ethereum network, and  Fabric version v1.3.0 \footnote{https://github.com/hyperledger/fabric-releases} for a local Hyperledger network.
%, and conduct experiments over the above four databases.
$\textit{TestRPC}$ is initialized with the default configuration, which is much like real Ethereum environment except that its block time for mining is set to be 0. This allows us to focus on the performance of SSE part on smart contract, irrespective of time-consuming mining process and complex network circumstances (\eg, broadcast latency, transaction mining delay) in Ethereum.

Table~\ref{tab:timecost_rpc} presents an overview of time costs for each phase over different datasets. In the setup phase, different from existing centralized SSE schemes where the data owner side dominates the efficiency, the time cost on smart contract is much higher than that on the data owner.
This is because storing $\textsf{EDB}$ in $\Pi$ is completed with thousands of transactions, with each transaction costing 4 seconds on average, while $\Pi^+$ needs about hundreds of transactions.
We also observe that $\Pi^+$ has a much higher efficiency in each step than $\Pi$. This again shows that the private blockchain leverages a faster consensus algorithm (\eg, PBFT vs. PoW), such that $\Pi^+$ inevitably enjoys a higher efficiency than $\Pi$ although they have the same structure of encrypted index.

\begin{figure}[t]
\centering
\subfigure[Search time per matching document in $\Pi$ and $\Pi^+$.]{\label{fig:searchtim_rpc}
\includegraphics[width=0.47\columnwidth]{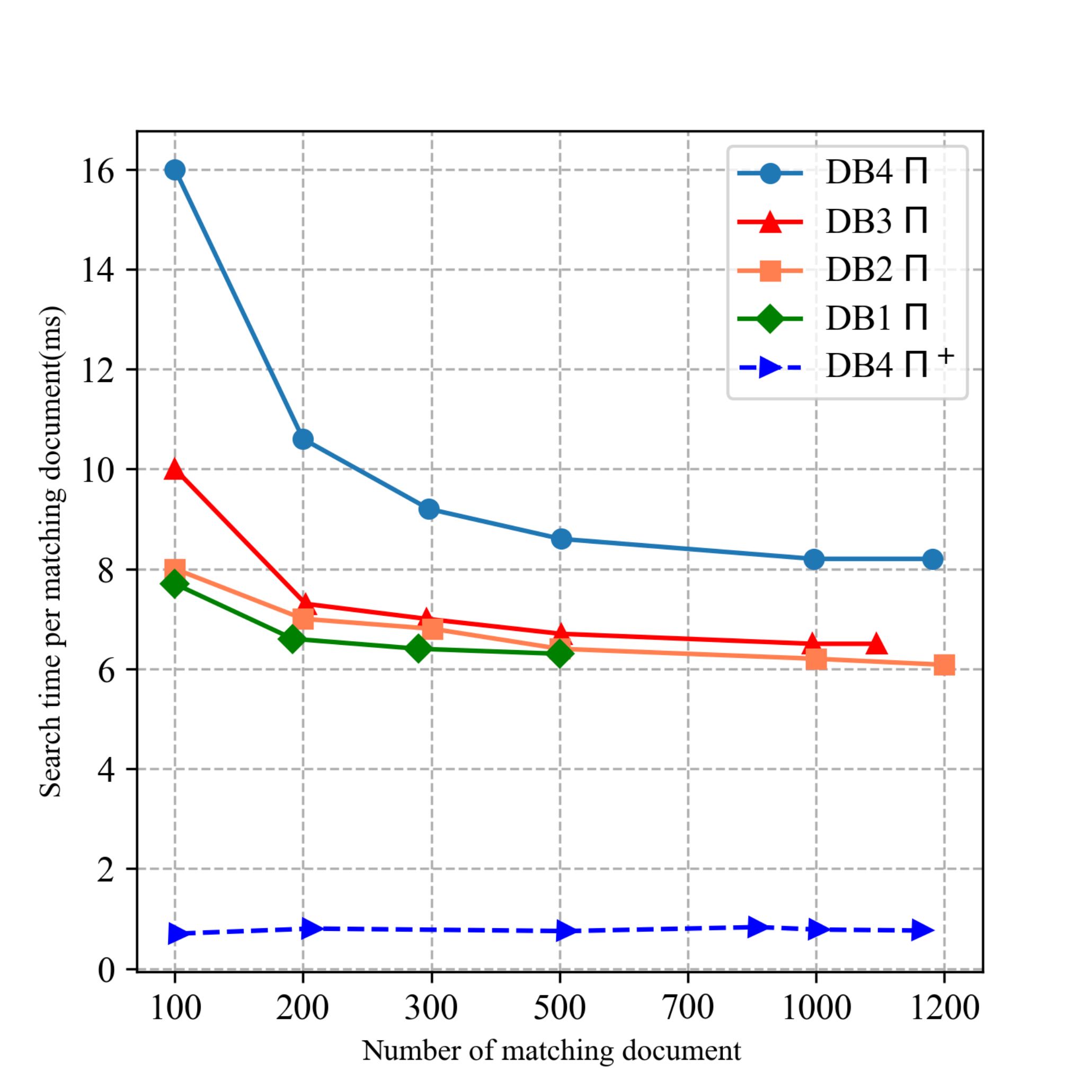}}
\subfigure[Update time vs. the number of added files in $\Pi^+$.]{\label{fig:updatetime_fabric}
\includegraphics[width=0.47\columnwidth]{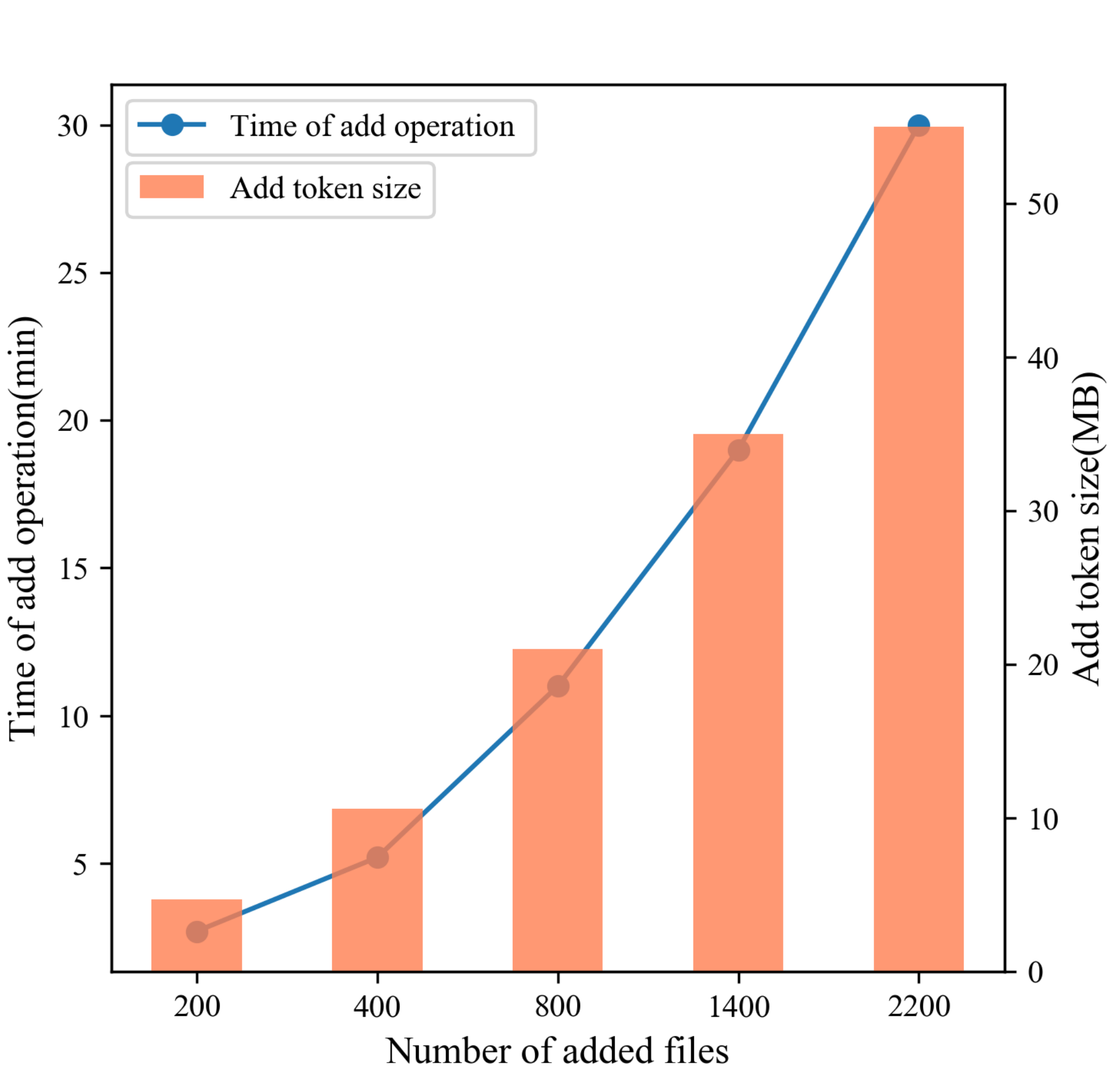}}
%\vspace{-2mm}
\caption{ Efficiency evaluations for $\Pi$ and $\Pi^+$.}
\vspace{-2mm}
\end{figure}

To show the core algorithm, Fig.~\ref{fig:searchtim_rpc} presents the search time per found document varying with the number of matching records.
Due to the high efficiency of $\Pi^+$, we only evaluate it with the largest dataset DB4.
We report average run times over 30 trials. For $\Pi$, the first thing we can notice is that a larger result set yields a lower search overhead (on a per matching document basis). We explain that by the constant cost of loading past mined blocks from disk into memory before each search runs.
%The more mined blocks, the longer it takes for loading.
This also explains our second observation: the larger the dataset, the slower the search algorithm is. A larger number of mined blocks leads to a longer time for loading. For $\Pi^+$, we not only see that it has a lower time cost than $\Pi$, but also conclude that the number of matching document has negligible impact on the search overhead.

\begin{figure}[t]
\centering
\subfigure[Search time vs. the number of matching documents]{\label{fig:searchtime_rinke}
\includegraphics[width=0.47\columnwidth]{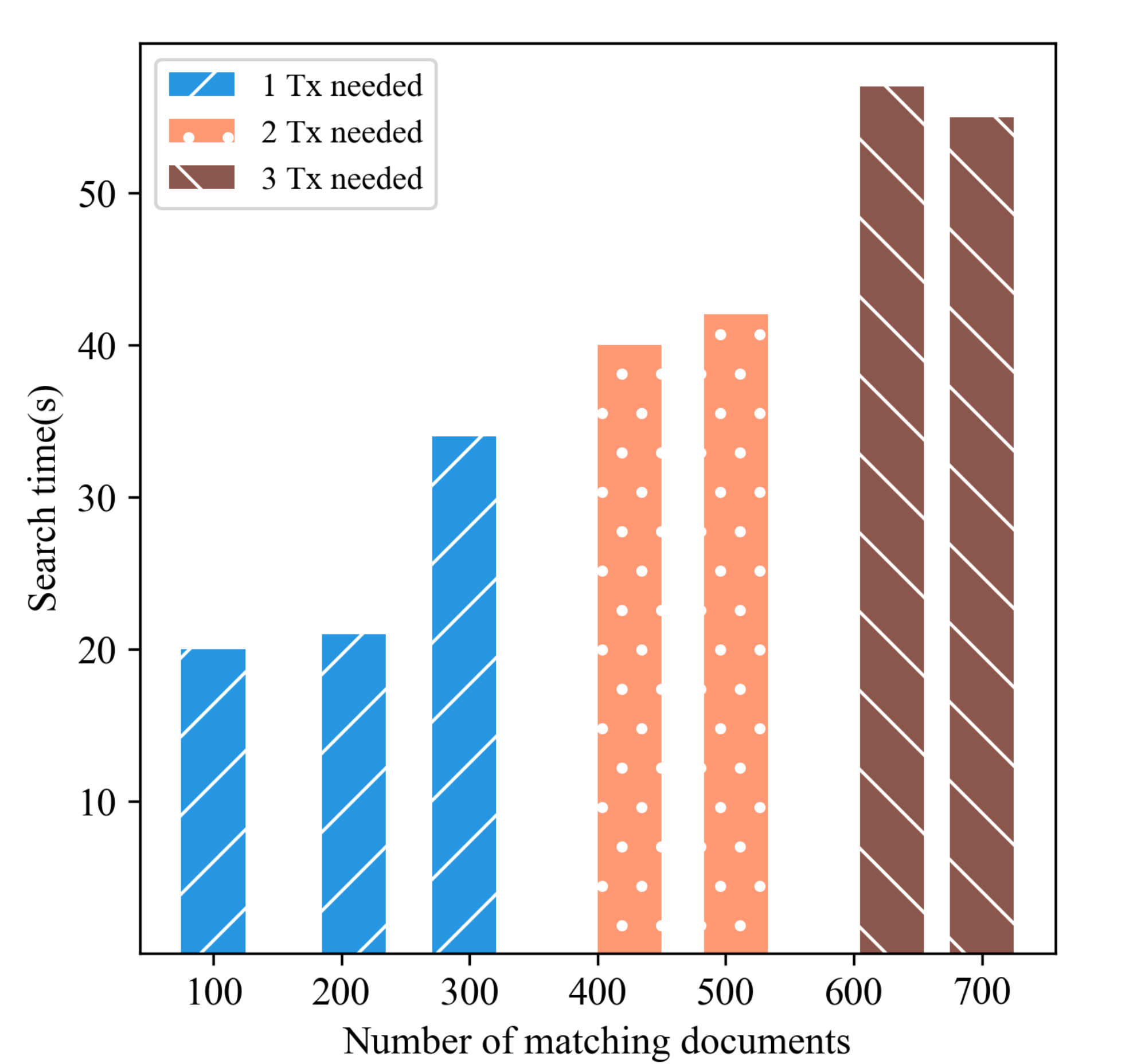}}
\subfigure[Update time vs. the number of transactions.]{\label{fig:updatetime_rinke}
\includegraphics[width=0.47\columnwidth]{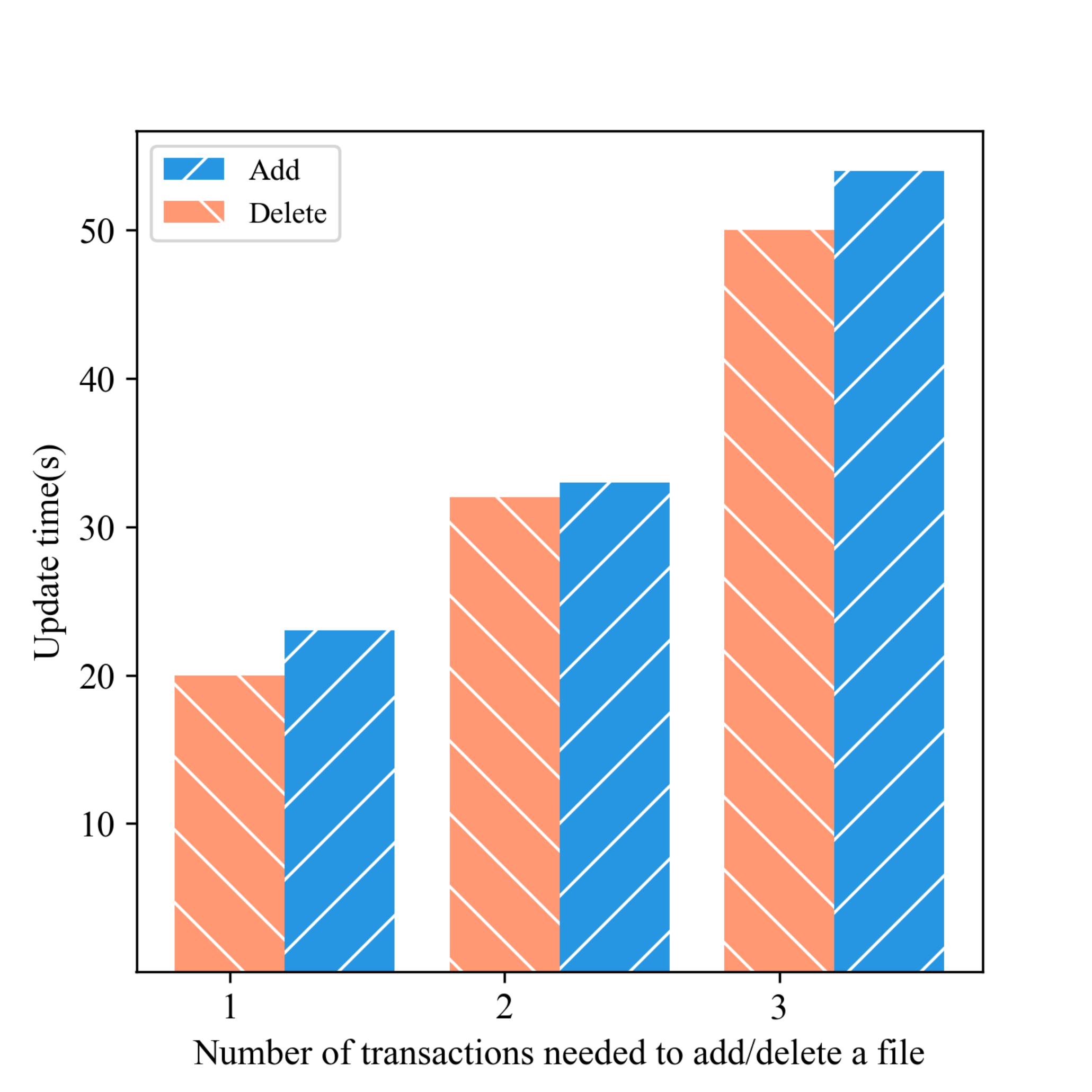}}
%\vspace{-2mm}
\caption{ Efficiency evaluations for $\Pi$ in Rinkeby.}
%\vspace{-2mm}
\end{figure}

Fig.~\ref{fig:updatetime_fabric} shows the update performance for $\Pi^+$ by varying the number of added files. Each added file includes 100 keyword/identifier pairs. We can see that adding about 2,200 files  costs less than half an hour.   $\Pi$ is not presented since we give a high-throughput experiment (\eg, adding hundreds of files) which is apparently not suitable for $\Pi$. The update experiments for $\Pi$  over a small dataset is shown in Fig.~\ref{fig:updatetime_rinke}.

To evaluate the performance of multi-user setting, Table~\ref{multiuser} presents the time costs of each algorithm described in Fig.~\ref{fig:multiuser}. We select the number of users in a large range to clearly demonstrate the efficiency. For the search process, we only present the additional time cost caused by $\textbf{Search}(\widetilde{ST}, \delta)$, without including the time cost of $\textbf{Search}(ST)$.
We can see that the time costs of \textbf{Setup} and \textbf{RevokeUser} increase with the number of users, and they have similar overheads. This is because revoking users in $\Pi^+$ needs to generate new secret keys and renewedly perform broadcast encryption in the same way with the setup. On the contrary, the other operations incur negligible time costs. Compared with the frequently executed searching, revoking users can be regarded as an one-time operation. Therefore, we emphasize that our multi-user design is still practical in real-world applications.

\begin{figure}[!t]
\centering
\includegraphics[width=0.8\columnwidth]{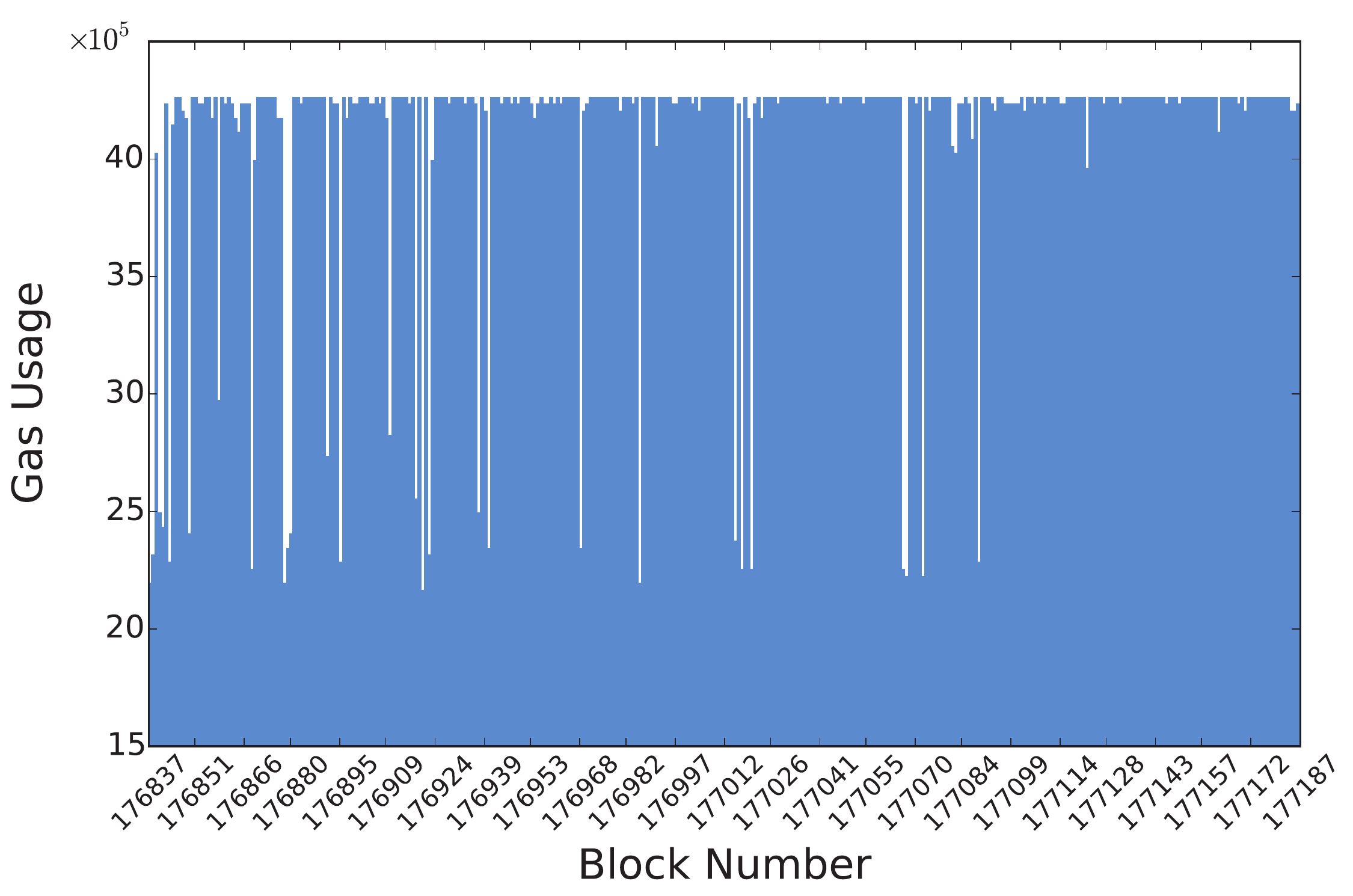}
%\vspace{-2mm}
\caption{Setup: Gas usage of each mined block in Rinkeby. }
\label{fig:gasusage}
%\vspace{-2mm}
\end{figure}

\subsection{Experiments on Official Test Network}

To show the practicability of decentralized SSE scheme, we deploy $\Pi$ to the official Ethereum test network \textit{Rinkeby}\footnote{https://www.rinkeby.io/} that mimics the real production network. Due to the limited balance, we only conduct experiments on the smallest database \textsf{DB1}. Our account and contract addresses in Rinkeby are
\begin{itemize}
  \item 0x7aef688b95a1bee573d464766b3a6c0470b9b57b.
  \item 0xecE97a98Da7f5DBECcb81E772dD04710e676Aa96.
\end{itemize}
To illustrate the impact of mining process on the efficiency, we record the block number of each transaction generated in our setup phase and the corresponding gas usage, as shown in Fig.~\ref{fig:gasusage}. In summary, it consists of 350 transactions, each of which is mined into one block with block number ranging from $176,837$ to $177,187$. The average block time for mining is $15s$, resulting in $88min$ to complete the entire setup phase.
This again explains why the time cost of setup is dominated by the smart contract, instead of the data owner like in existing centralized SSE schemes.
Besides, the average gas usage for a transaction is $4,201,232$. Currently 1 gas costs about $1.8\times 10^{-8}$ Ether, at the exchange rate of 89 USD at the time of writing.  So each transaction costs about $0.076$ Ether (or $6.7$ USD).

Fig.~\ref{fig:searchtime_rinke} shows the total time needed to perform a search, given a search token (we neglect the cost of generating a search token since it is a small constant in microseconds). Each point is the mean of 10 executions. It clearly demonstrates the performance bottleneck of decentralized SSE. To be specific, we can see that the search time grows with the increase of the number of matching documents. But the sharp growth lies in the increase of the transaction number needed to complete the search step. It indicates that the time cost of mining each transaction dominates the overhead of each search. On the contrary, search algorithm has a faint impact on the efficiency. Generally speaking, the time cost of the mining process is dynamically adjustable. When the blockchain environment scales to allow a higher gas limitation or a faster mining process, our search efficiency increases as well.

A similar situation occurs in Fig.~\ref{fig:updatetime_rinke} which describes time costs varying with the number of transactions needed to add/delete a file. By choosing different sizes of files, we have update completed with different numbers of transactions. It again shows that the mining process of each transaction is the dominant factor on the efficiency.

\section{Future Work}

\subsection{Hardening  Security with Trusted Processor}
Trusted processor  is one of emerging security technologies that protects the private information through hardware-assisted trusted execution environment.
It can protect the integrity and confidentiality of private data from other applications and privileged system software such as the operating system, hypervisor, and firmware, and has been widely used to provide privacy guarantee for various jobs, like Tor network\cite{kim2017enhancing} or system log processing~\cite{karande2017sgx}, \etc Although $\Pi^+$ and $\Pi$ are designed to secure the private data, some information leakage still exists such as the search pattern and access pattern. In light of this, integrating trusted processor with blockchain is a promising approach to address this issue.  Prior works have explored the potential of applying trusted processors to the encrypted search~\cite{mishra2018oblix,fuhry2017hardidx,hu2019towards}, but how to support blockchain-based decentralized encrypted search is still a challenging problem.

\subsection{Improving Efficiency with Sharding}
Sharding is an important technique to improve the efficiency and scalability of blockchain networks. It generally partitions a large blockchain network into separate subsets (\ie, shards), each of which deals with a disjoint set of transactions and runs an intra consensus protocol independently~\cite{luu2016secure,kokoris2018omniledger}. It is obvious that building our schemes of $\Pi$ and $\Pi^+$ atop of sharded blockchains benefits a lot for improving efficiency. Besides, a tailor-made search index is desired that caters to the sharded structure of blockchain. Parallel execution of search operations among shards can also improve efficiency greatly. However, how to design such a customized encrypted search index still remains unclear.

\section{Conclusion}
Traditional searchable symmetric encryption relies on a central server to manipulate search jobs.
In this work, we resort to public and private blockchain technologies and construct two decentralized SSE schemes aiming at addressing malicious adversary. Different from existing verifiable SSE schemes, our
search results are correct and immutable, and no verifications are needed on the data owner side. Our framework can be applied to other SSE schemes with complex queries.
 Finally, we conduct extensive experiments in both locally simulated and official test networks to demonstrate the practicability of decentralized SSE schemes.

\appendix[Confidentiality Proof]\label{sec:append}

We first restate the security claim for $\Pi$.
\begin{theorem}
If $G$ and $F$ are pseudo-random, define $\mathcal{L}=(\mathcal{L}_1$, $\mathcal{L}_2$, $\mathcal{L}_3)$, then our scheme $\Pi$ is $\mathcal{L}$-secure against non-adaptive attacks.
\end{theorem}
\begin{pf}
We describe a polynomial-time simulator $\mathcal{S}$ such that for any PPT adversary $\mathcal{A}$, the outputs of $\mathbf{Real}^{\Pi}_{\mathcal{A}}(\lambda)$ and $\mathbf{Ideal}^{\Pi}_{\mathcal{A}, \mathcal{S}}(\lambda)$ are computationally indistinguishable.

To prove non-adaptive security, the simulator $\mathcal{S}$ must be given all the leakages before simulating the view of the adversary, which includes the encrypted database ($\gamma$, $\gamma^A$ and $\mathsf{ID}_{\mathsf{del}}$) and the messages sent by the data owner.

The simulator iterates over the queries, it chooses the keys $\widetilde{K_1},\widetilde{K_2},\widetilde{K_1^A},\widetilde{K_2^A},\widetilde{K_1^D}$ for each search at random with repetitions specified by the search pattern. Then it simulates the initial $\mathsf{EDB}$ as follows. For all file $\mathsf{id}$'s associated with each search keyword $w$ (\ie, $\mathsf{id} \in \mathsf{DB}(w)$), $\mathcal{S}$ computes $l$, $d$ and $r$ as specified in the real $\mathsf{Setup}$ (using $\widetilde{K_1}$ and $\widetilde{K_2}$ as $K_1$ and $K_2$), adds each pair $(l,d,r)$ to a list $L$, and then adds random pairs to $L$ (still maintained in lexicographic order) until it has $\sum_{w\in \mathsf{W}}\lceil \frac{|\mathsf{DB}(w)|}{p} \rceil$ total elements, and finally creates a dictionary $\widetilde{\gamma}$. The simulator outputs the simulated $\widetilde{\gamma}$ and the simulated transcript $(\widetilde{K_1},\widetilde{K_2},\widetilde{K_1^A},\widetilde{K_2^A},\widetilde{K_1^D})$ for each search query. Note that $c$ and $\textsf{step}$ are public system parameters and deterministically computable from the state information, which do not need to be simulated.

Next, to simulate add update queries, that is, the simulator $\mathcal{S}$ needs to simulate the message sent by the client, which contains multiple $label/data/randomness/id_{\textsf{del}}$ tuples. The simulator $\mathcal{S}$ must determine whether each tuple sent is generated at random or should be computed with one of the keys used for a search query transcript. Intuitively, this can be done by leveraging both the add pattern $\mathsf{AP}(w,Q,\mathsf{ID})$ and $\mathsf{ap}(\mathsf{id},w,Q)$ leakages which include the $\mathsf{id}$ to encrypt when the add updates contain a keyword that was previously searched. The simulator can further simulate $\widetilde{\gamma^A}$ based on all the messages sent by the client and the delete patterns. Note that the message sent back to the client is deterministically computable from the state information, which does not need to be simulated.

To simulate delete update queries, the simulator $\mathcal{S}$ needs to simulate the message sent by the client like add. Thus by using the deletion patterns $\mathsf{DP}(w,Q,\mathsf{ID})$ and $\mathsf{dp}(\mathsf{id},w,Q)$ leakages, the simulator can simulate the corresponding message in a similar way. Finally, the simulator can simulate $\widetilde{\mathsf{ID}_{\mathsf{del}}}$ based on all the messages sent by the client and the add patterns.

In summary, the theorem follows from the pseudo-randomness of $F$ and $G$.
\end{pf}

\ifCLASSOPTIONcaptionsoff
  \newpage
\fi

\bibliographystyle{IEEEtran}
\bibliography{ref}

\vspace{-15mm}
\begin{IEEEbiography}[{\includegraphics[width=1in,height=1.25in,clip,keepaspectratio]{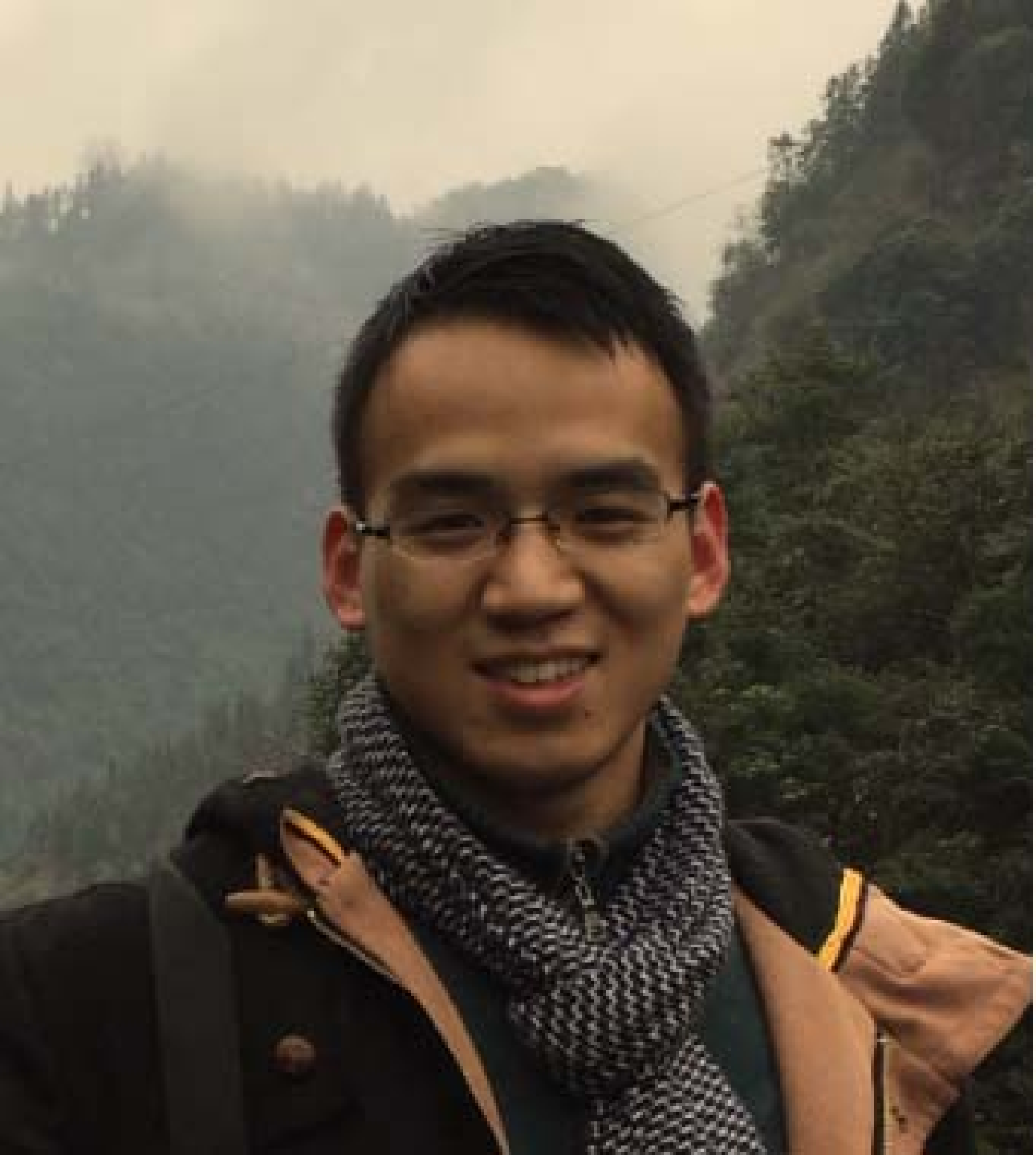}}]{Shengshan Hu}
 received the B.E. degree from Wuhan University, China, in 2014, in Computer Science and Technology. He is currently a Ph.D. candidate in the School of Cyber Science and Engineering, Wuhan University. His research interest focuses on secure outsourcing of computations.
\end{IEEEbiography}

\vspace{-5mm}
\begin{IEEEbiography}[{\includegraphics[width=1in,height=1.25in,clip,keepaspectratio]{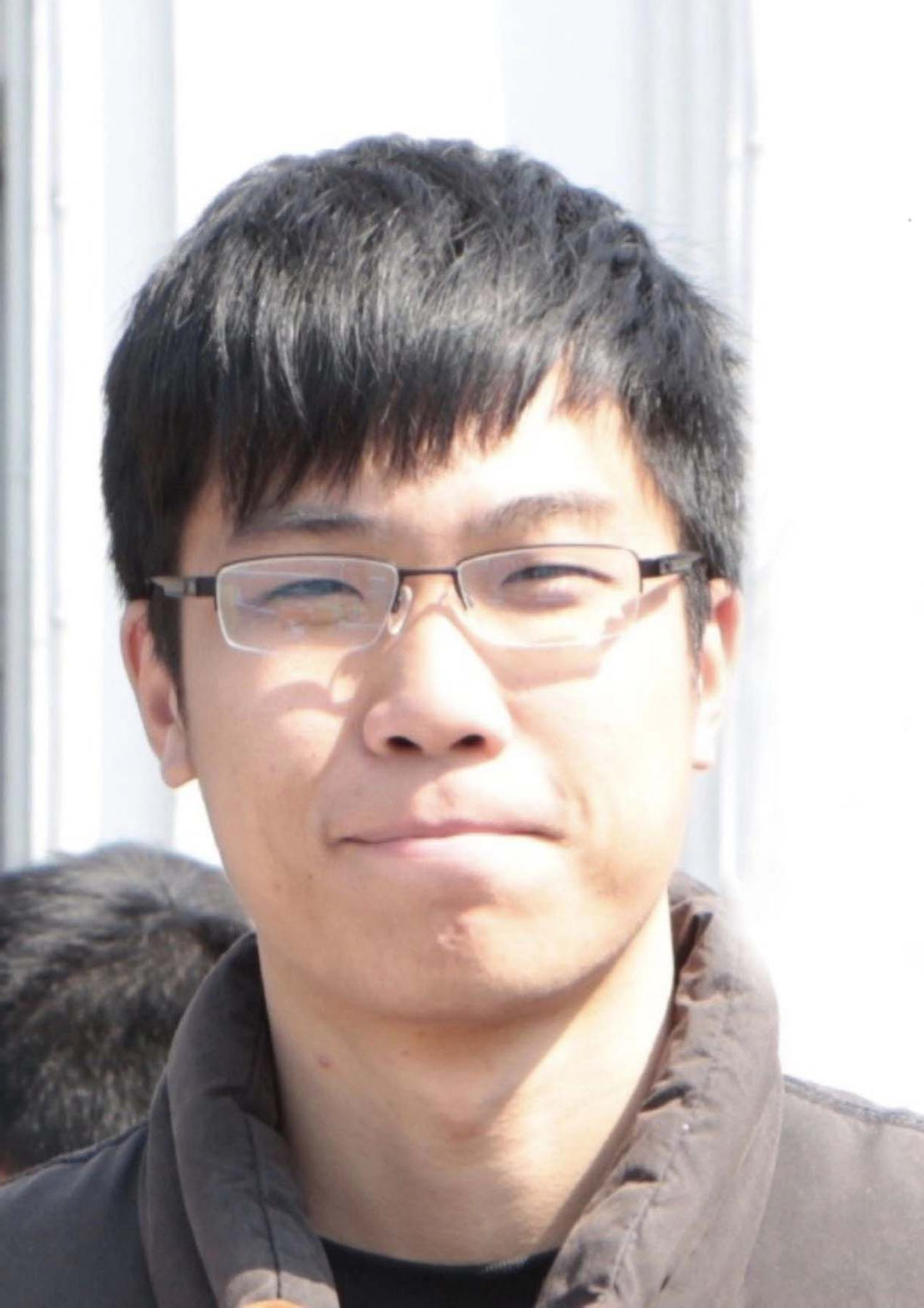}}]{Chengjun Cai} received the BS degree in computer science and technology from Jinan University, in 2016. He is working toward the PhD degree at the City University of Hong Kong. He was a research assistant with the City University of Hong Kong. His research interests include distributed system security and privacy-enhancing technologies. He is a student member of the IEEE.
\end{IEEEbiography}

%\vspace{-5mm}
%\begin{IEEEbiography}[{\includegraphics[width=1in,height=1.15in,clip,keepaspectratio]{Zhan_Qin}}]{Qin Zhan} is an Assistant Professor in the Department of Electrical and Computer Engineering at The University of Texas at San Antonio. His research interests include secure computation outsourcing, privacy-preserving data collection, sharing and publication, cybersecurity of smart grid control and communication system, and cyberphysical security for smart devices with the current
%focus on exploring and improving the security and privacy assurance on cloud computing.
%
%\end{IEEEbiography}

\vspace{-5mm}
\begin{IEEEbiography}[{\includegraphics[width=1in,height=1.25in,clip,keepaspectratio]{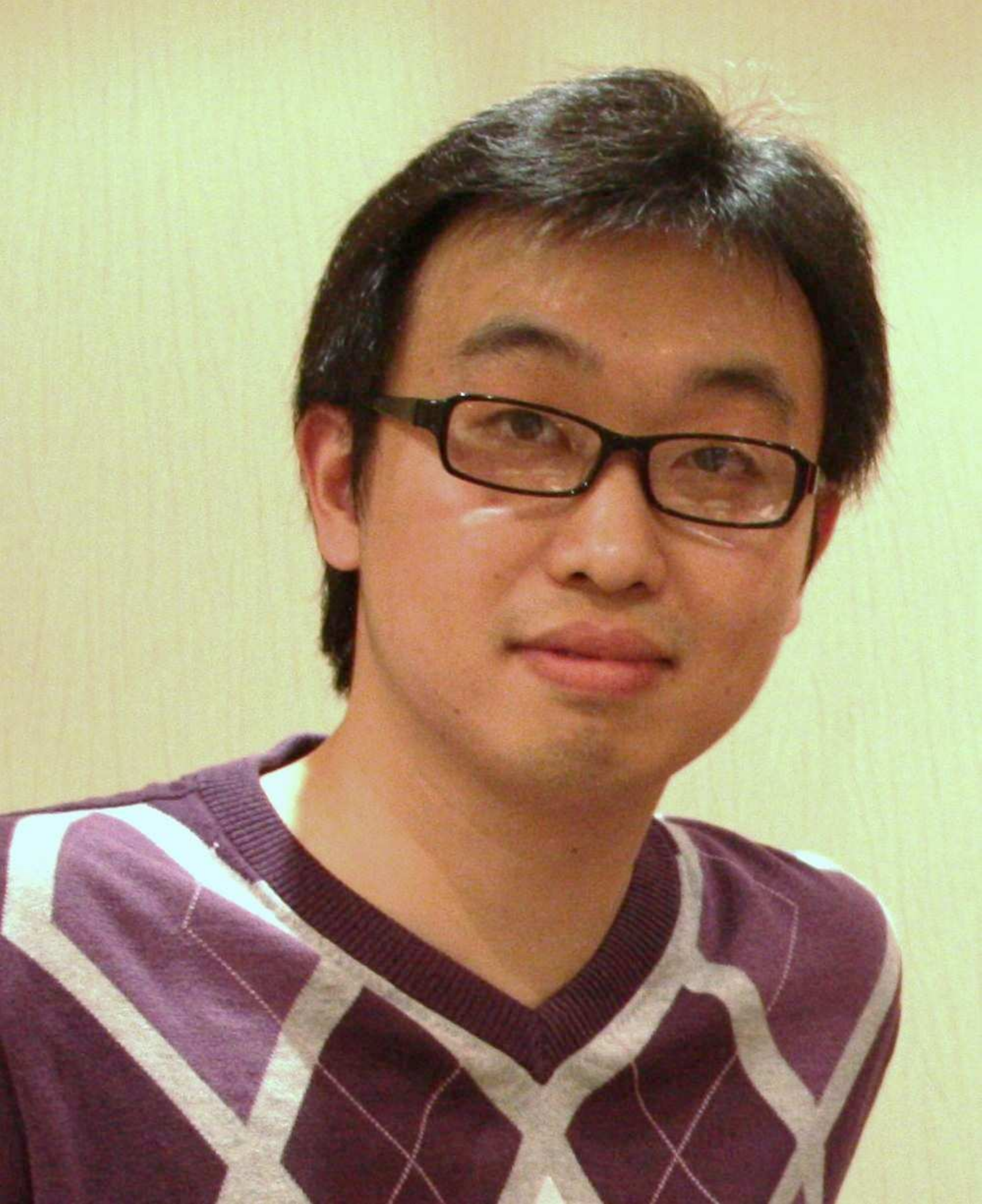}}]{Qian Wang} is a Professor with the School of Cyber Science and Engineering, Wuhan University. He received the Ph.D. degree from Illinois Institute of Technology, USA. His research interests include AI security, data storage, search and computation outsourcing security and privacy, wireless systems security, big data security and privacy, and applied cryptography etc. Qian received National Science Fund for Excellent Young Scholars of China in 2018. He is also an expert under National ``1000 Young Talents Program'' of China. He is a recipient of the 2018 IEEE TCSC Award for Excellence in Scalable Computing (Early Career Researcher), and the 2016 IEEE Asia-Pacific Outstanding Young Researcher Award. He is also a co-recipient of several Best Paper and Best Student Paper Awards from IEEE ICDCS'17, IEEE TrustCom'16, WAIM'14, and IEEE ICNP'11 etc. He serves as Associate Editors for IEEE Transactions on Dependable and Secure Computing (TDSC) and IEEE Transactions on Information Forensics and Security (TIFS). He is a Member of the IEEE and a Member of the ACM.
\end{IEEEbiography}

\vspace{-5mm}
\begin{IEEEbiography}[{\includegraphics[width=1in,height=1.25in,clip,keepaspectratio]{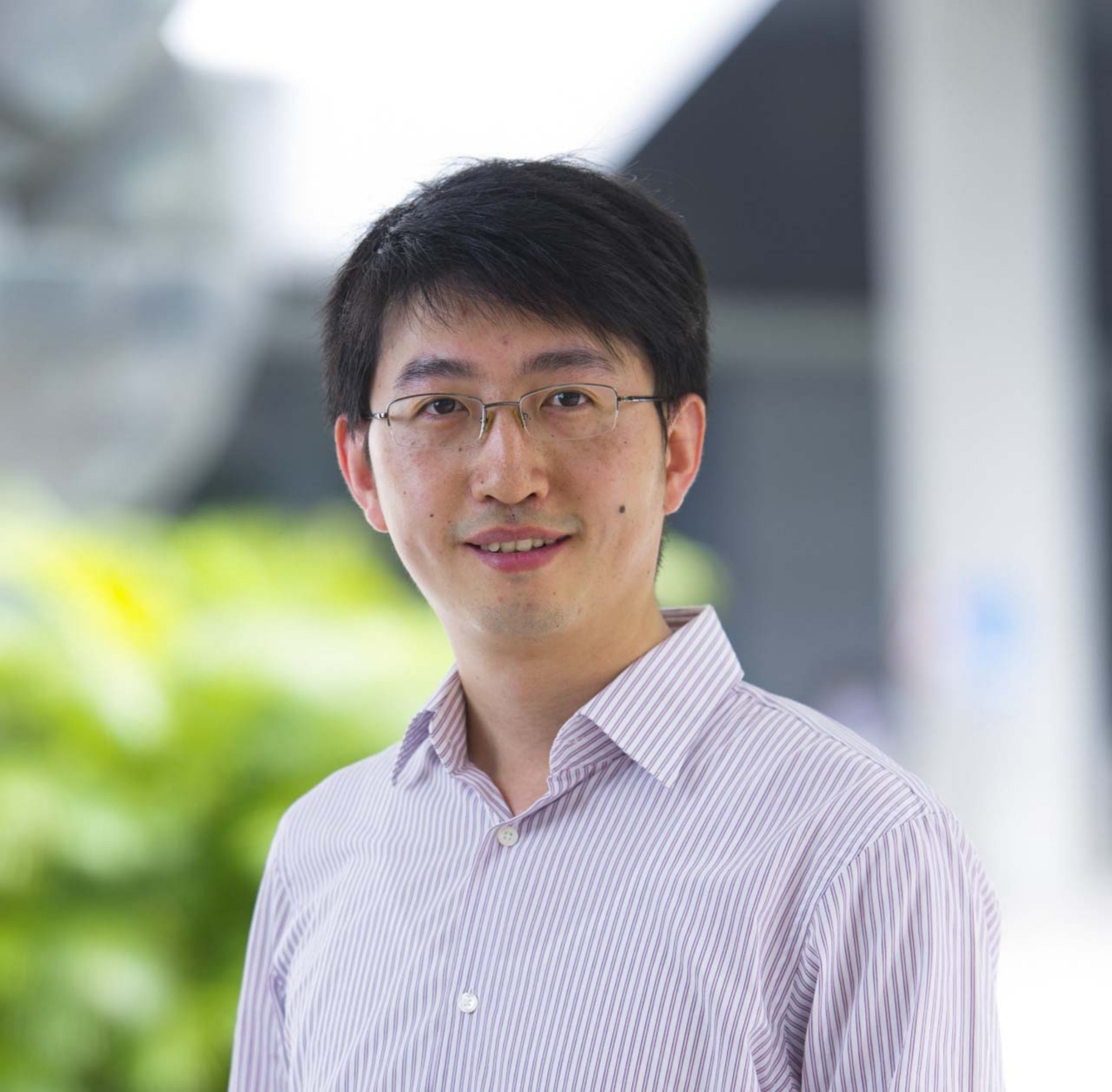}}]{Cong Wang}
has been an Assistant Professor at the Department of Computer Science, City University of Hong Kong, since the Summer of 2012. He received his PhD in the Electrical and Computer Engineering from Illinois Institute of Technology, USA, and M.Eng and B.Eng from Wuhan University, China. His current research interests include data and computation outsourcing security in the context of cloud computing, network security in emerging Internet architecture, multimedia security and its applications, and privacy-enhancing technologies in the context of big data and IoT. He received The President's Awards 2016 at City University of Hong Kong. He was the co-recipient of the Best Student Paper Award of IEEE ICDCS 2017, and the Best Paper Award of IEEE MSN 2015 and CHINACOM 2009. His research has been supported by multiple government research fund agencies, including National Natural Science Foundation of China, Hong Kong Research Grants Council, and Hong Kong Innovation and Technology Commission. He has been serving as the TPC co-chairs for a number of IEEE conferences/workshops. He is a member of IEEE and ACM.
\end{IEEEbiography}

\vspace{-5mm}
\begin{IEEEbiography}[{\includegraphics[width=1in,height=1.15in,clip,keepaspectratio]{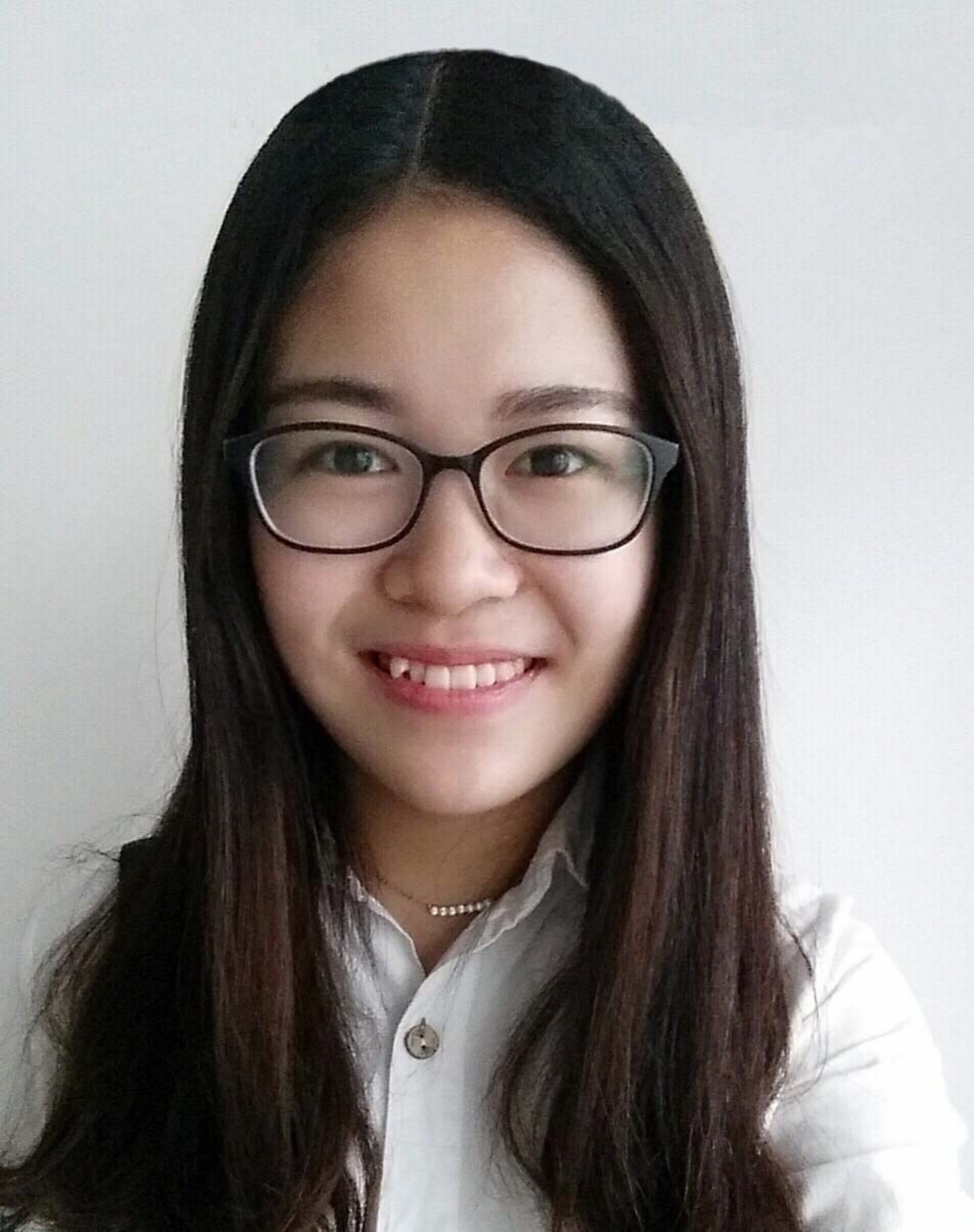}}]{Minghui Li}
received the B.S. degree in Information Security from Wuhan University, China, in
2016. She is currently a graduate student in the School of Cyber Science and Engineering, Wuhan University, China. Her research interest focuses on privacy-preserving machine learning.
\end{IEEEbiography}

\begin{IEEEbiography}[{\includegraphics[width=1in,height=1.25in,clip,keepaspectratio]{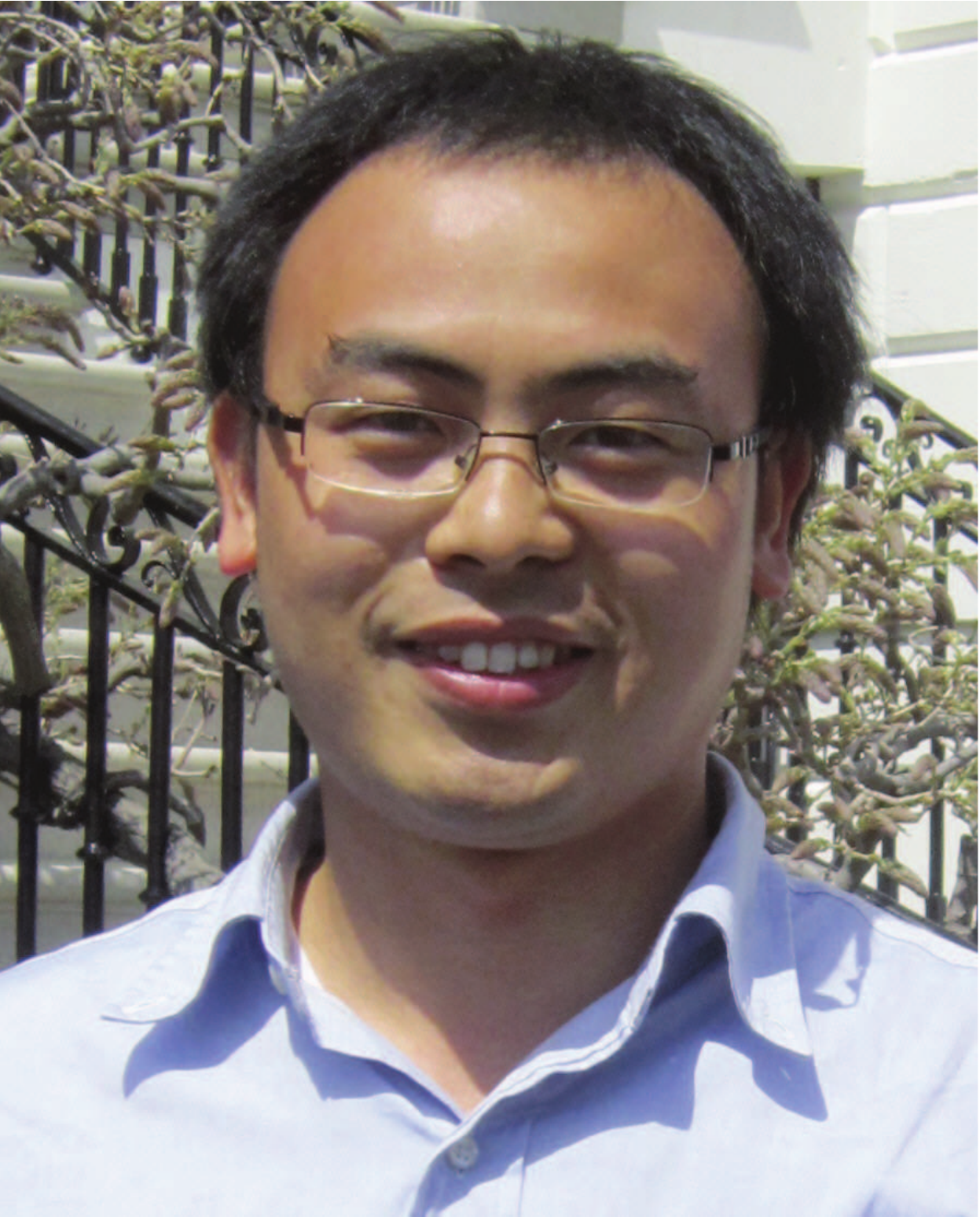}}]{Zhibo Wang}
received the B.E. degree in Automation from Zhejiang University, China, in 2007, and his Ph.D degree in Electrical Engineering and Computer Science from University of Tennessee, Knoxville, in 2014. He is currently a Professor with the School of Cyber Science and Engineering, Wuhan University, China. His currently research interests include wireless sensor networks and mobile sensing systems. He is a Senior Member of the IEEE and a Member of the ACM.
\end{IEEEbiography}

\vspace{-5mm}
\begin{IEEEbiography}[{\includegraphics[width=1in,height=1.25in,clip,keepaspectratio]{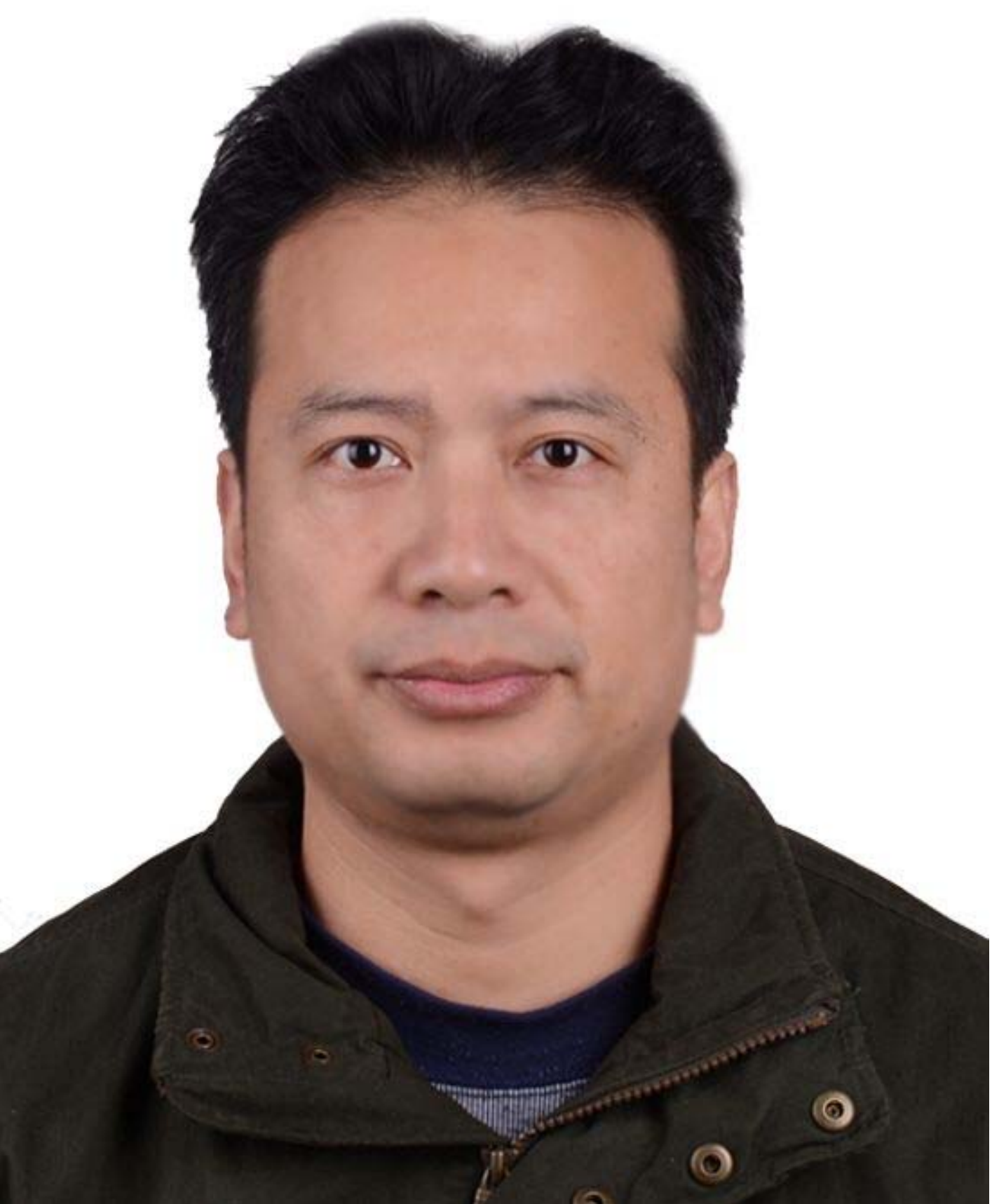}}]
{Dengpan Ye} received the B.S.degree in automatic control from SCUT in 1996 and the Ph.D. degree from NJUST in 2005. He was a Post-Doctoral Fellow in information system with the School of Singapore Management University. Since 2012, he has been a Professor with the School of Cyber Science and Engineering, Wuhan University. His research interests include machine learning and multimedia security. He has authored or coauthored over 50 refereed journal and conference papers.
\end{IEEEbiography}

\end{document}